\documentclass[pdflatex,sn-nature]{sn-jnl}% Style for submissions to Nature Portfolio journals
\usepackage{graphicx}%
\usepackage{multirow}%
\usepackage{gensymb}%
\usepackage{amsmath,amssymb,amsfonts}%
\usepackage{amsthm}%
\usepackage{mathrsfs}%
\usepackage[title]{appendix}%
\usepackage{xcolor}%
\usepackage{textcomp}%
\usepackage{manyfoot}%
\usepackage{booktabs}%
\usepackage{algorithm}%
\usepackage{algorithmicx}%
\usepackage{algpseudocode}%
\usepackage{listings}%
\usepackage{hyperref}

\makeatletter
\gdef\orcidlogo{%
	\includegraphics[height=1.6ex]{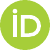}%
}
\makeatother

\theoremstyle{thmstyleone}%
\theoremstyle{thmstyletwo}%

\theoremstyle{thmstylethree}%

\def \src {1E~1547.0$-$5408}

\def\apj{ApJ}
\def\apjl{ApJL}
\def \apspr{Astrophysics Space Physics Research}
\def\aplett{Astrophysical Letters}
\def\aap{A\&A}
\def\araa{Annual Review of Astronomy and Astrophysics}
\def\aj{AJ}
\def\nat{Nature}
\def\pasa{PASA}
\def\prd{Physical Review D}
\def\prl{Physical Review Letters}
\def\pasp{Publications of the Astronomical Society of the Pacific}
\def\mnras{MNRAS}
\newcommand{\physrep}{Physics Reports}

\begin{document}
%TC:ignore
%\linenumbers
\title[Highly polarized thermal X-ray emission from \src]{\bf Vacuum birefringence and the polarized X-ray emission from a radio magnetar}

%Exceptional polarization of thermal X-rays from a radio magnetar


\author*[1]{\fnm{Rachael~E.} \sur{Stewart}\,\orcid{https://orcid.org/0000-0002-0254-5915}}\email{raestewart@gwu.edu}
\equalcont{These authors contributed equally to this work.}

\author*[2]{\fnm{Hoa} \sur{Dinh Thi}\,\orcid{https://orcid.org/0000-0001-9268-5577}}\email{hoa.dinh@rice.edu}
\equalcont{These authors contributed equally to this work.}

\author*[3,4]{\fnm{George} \sur{Younes}\,\orcid{https://orcid.org/0000-0002-7991-028X}}\email{george.a.younes@nasa.gov}

\author*[5]{\fnm{Marcus~E.} \sur{Lower}\,\orcid{https://orcid.org/0000-0001-9208-0009}}\email{mlower@swin.edu.au}

\author*[2]{\fnm{Matthew~G.} \sur{Baring}\,\orcid{https://orcid.org/0000-0003-4433-1365}}\email{baring@rice.edu}

\author[6]{\fnm{Michela} \sur{Negro}\,\orcid{https://orcid.org/0000-0002-6548-5622}}

\author[7]{\fnm{Fernando} \sur{Camilo}\,\orcid{https://orcid.org/0000-0002-1873-3718}}

\author[8]{\fnm{Joel~B.} \sur{Coley}\,\orcid{https://orcid.org/0000-0001-7532-8359}}

\author[9]{\fnm{Teruaki} \sur{Enoto}\,\orcid{https://orcid.org/0000-0003-1244-3100}}

\author[10]{\fnm{Alice~K.} \sur{Harding}\,\orcid{https://orcid.org/0000-0001-6119-859X}}

\author[11]{\fnm{Wynn~C.~G.} \sur{Ho}\,\orcid{https://orcid.org/0000-0002-6089-6836}}

\author[12]{\fnm{Chin-Ping} \sur{Hu}\,\orcid{https://orcid.org/0000-0001-8551-2002}}

\author[13]{\fnm{Philip} \sur{Kaaret}\,\orcid{https://orcid.org/0000-0002-3638-0637}}

\author[14]{\fnm{Paul} \sur{Scholz}\,\orcid{https://orcid.org/0000-0002-7374-7119}}

\author[1] {\fnm{Alex} \sur{Van Kooten}\,\orcid{https://orcid.org/0000-0002-3905-4853}}

\author[15,3]{\fnm{Zorawar} \sur{Wadiasingh}\,\orcid{https://orcid.org/0000-0002-9249-0515}}

\affil[1]{\orgdiv{Department of Physics}, \orgname{George Washington University}, \orgaddress{\street{725 21st Street NW}, \city{Washington DC}, \postcode{20052}, \state{District of Columbia}, \country{USA}}}

\affil[2]{\orgdiv{Department of Physics and Astronomy}, \orgname{Rice University}, \orgaddress{\street{6100 Main Street}, \city{Houston}, \postcode{77251}, \state{Texas}, \country{USA}}}

\affil[3]{\orgdiv{Astrophysics Science Division}, \orgname{NASA Goddard Space Flight Center}, \orgaddress{\street{8800 Greenbelt Road}, \city{Greenbelt}, \postcode{20771}, \state{Maryland}, \country{USA}}}

\affil[4]{\orgdiv{Center for Space Sciences and Technology}, \orgname{University of Maryland Baltimore County}, \orgaddress{\street{1000 Hilltop Cir}, \city{Baltimore County}, \postcode{21250}, \state{Maryland}, \country{USA}}}

\affil[5]{\orgdiv{Centre for Astrophysics and Supercomputing}, \orgname{Swinburne University of Technology}, \orgaddress{\street{PO Box 218}, \city{Hawthorn}, \postcode{3122}, \state{Victoria}, \country{Australia}}}

\affil[6]{\orgdiv{Department of Physics and Astronomy}, \orgname{Louisiana State University}, \orgaddress{\street{202 Nicholson Hall}, \city{Baton Rouge}, \postcode{70803}, \state{Louisiana}, \country{USA}}}

%\affil[7]{\orgdiv{Department}, \orgname{Organization}, \orgaddress{\street{Street}, \city{City}, \postcode{610101}, \state{State}, \country{Country}}}

\affil[7]{\orgname{South African Radio Astronomy Observatory}, \orgaddress{\street{Liesbeek House, River Park}, \city{Cape Town}, \postcode{7700}, \country{South Africa}}}

\affil[8]{Department of Physics and Astronomy, Howard University, Washington, DC 20059, USA; CRESST/Mail Code 661, Astroparticle Physics Laboratory, NASA Goddard Space Flight Center, Greenbelt, MD 20771, USA}

\affil[9]{\orgdiv{Department of Physics}, \orgname{Kyoto University}, \orgaddress{\city{Kyoto}, \country{Japan}}}

\affil[10]{\orgdiv{Theoretical Division}, \orgname{Los Alamos National Laboratory}, \orgaddress{\city{Los Alamos}, \postcode{87545}, \state{New Mexico}, \country{USA}}}

\affil[11]{\orgdiv{Department of Physics and Astronomy}, \orgname{Haverford College}, \orgaddress{\street{370 Lancaster Avenue}, \city{Haverford}, \postcode{19041}, \state{Pennsylvania}, \country{USA}}}

\affil[12]{\orgdiv{Department of Physics}, \orgname{National Changhua University of Education}, \orgaddress{\street{No. 1, Jinde Rd.}, \city{Changhua}, \postcode{500207}, \country{Taiwan}}}

\affil[13]{\orgname{NASA Marshall Space Flight Center}, \orgaddress{ \city{Huntsville}, \postcode{35812}, \state{AL}, \country{USA}}}

\affil[14]{\orgdiv{Department of Physics and Astronomy}, \orgname{York University}, \orgaddress{\street{4700 Keele Street}, \city{Toronto}, \postcode{MJ3 1P3}, \state{ON}, \country{Canada}}}

\affil[15]{\orgname{Department of Astronomy, University of Maryland}, \orgaddress{ \city{College Park}, \postcode{20742}, \state{MD}, \country{USA}}}


%\abstract{}
\maketitle
\centerline{ \bf Summary paragraph}

\vspace{4pt}

{Magnetars are isolated neutron stars with exceptionally strong surface fields exceeding $10^{14}$~G \cite{kouv98Natur:1806}. Their bright X-ray emission probes physical regimes in which quantum electrodynamic (QED) influences radiation propagation \cite{Lai-Ho-2003PhRvL,vanadelsberg09MNRAS,taverna20MNRAS}. Strong magnetic fields induce polarization-dependent refractive indices in the vacuum \cite{Heisenberg-1936-ZPhy,Schwinger-1951-PhRv}; such vacuum birefringence (VB) remains a long-standing but unconfirmed prediction of QED. Here, we report phase- and energy-resolved polarization measurements of the radio-emitting magnetar \src\ obtained by coordinating X-ray and radio observations from the Imaging X-ray Polarimetry Explorer (IXPE), the Neutron Star Interior Composition ExploreR (NICER), and the Parkes/Murriyang observatory. We detect large polarization degrees (PD) in the thermally-dominant soft X-ray band, reaching phase-averaged values of 65\% at 2 keV before substantially decreasing between 2–4 keV. At certain rotational phases, the 2–3 keV PD rises to nearly 80\% while remaining high ($\gtrsim 40\%$) throughout the radio beam crossing. The phase-dependent X-ray and radio polarization angles are both consistent with the rotating vector model, suggesting that the emission geometries track the star's large-scale magnetic field. Collectively, these characteristics challenge standard surface emission models using non-refractive propagation of light to infinity.  VB-governed magnetospheric propagation can naturally explain the X-ray polarization signals. Our results represent a significant advance in probing this hallmark prediction of QED, opening a new cosmic window into superstrong-field quantum physics, thereby motivating further observational and theoretical studies concentrating on this domain.}

\newpage

Magnetars \cite{duncan92ApJmagnetars, paczynski92AcA:magnetars, kouv98Natur:1806}, a class of isolated neutron stars (NS), possess the strongest known magnetic fields in the observable universe (B$_{\text{surface}}\sim 10^{14-15}$ G), surpassing the typical field strength of young, rotation-powered pulsars by 1--3 orders of magnitude. The decay of these magnetic fields powers persistent, high-luminosity X-ray emission, which typically peaks between 0.5~keV and 10~keV and reaches $L_X \gtrsim 10^{35}$ erg s$^{-1}$ (see e.g. \cite{KaspiBeloborodov2017} for a review). With magnetic field strengths that far exceed the quantum electrodynamic (QED) critical field $B_{\text{cr}}=4.4 \times 10^{13}$ G, magnetars are widely recognized as promising astrophysical laboratories for testing exotic, non-linear QED processes in the strong B-field regime \cite{harding06RPPh,taverna24Galax}.

One such prediction is vacuum birefringence (VB), wherein intense magnetic fields -- comparable to those of magnetars -- substantially modify the refractive indices of photon polarization eigenmodes and render them disparate: different polarizations propagate at different speeds \cite{Adler-1971-AnPhy,Tsai-1975-PhRvD,Pavlov&Gnedin1984}.  This property is a direct consequence of the polarization of the quantum vacuum, and its constituent virtual electron-positron pairs, by the electromagnetic field \cite{Heisenberg-1936-ZPhy,Schwinger-1951-PhRv}, and remains an unverified prediction of QED. Indirect evidence consistent with the action of VB has been reported both in laboratory contexts \cite{2021STAR} and in astrophysical settings \cite{ho03ApJ5991293H,Mignani2017,taverna22Sci}, while direct measurements in either of these settings have remained challenging \cite{ho03ApJ5991293H, ho03MNRAS338233H, Ejlli2020,lai23PNAS,taverna24Galax}. Together, these complementary approaches provide important progress toward testing this fundamental physics prediction and offer promising avenues for further advances.

The influence of VB in magnetar magnetospheres is expected to strongly enhance the polarization signature of soft X-ray photons emerging from their surfaces. These photons are predicted to undergo differential phase shifts depending on their two polarization modes -- ordinary (O mode) and extraordinary (X mode). If {\bf k} represents a photon's momentum vector, the O mode ($\parallel$) has an electric field vector approximately in the {\bf k} - {\bf B} plane, and the X mode ($\perp$) has its electric field perpendicular to this plane.
VB decouples the polarization eigenmodes out to altitudes of $\sim 30 - 300$ stellar radii \cite{Heyl2003}, leading to a tight correlation of their polarization vectors with the magnetospheric field morphology. The result is that VB enhances the net linear polarization of surface emission, potentially producing polarization degrees (PDs) approaching 100\% under ideal viewing geometries \cite{Heyl-2000-MNRAS, Heyl2003, vanadelsberg09MNRAS}. Inside magnetar atmospheres, the combined influence of vacuum and plasma birefringence allows photons to convert between polarization modes \cite{Lai-Ho-2003PhRvL, lai23PNAS}. This mode conversion is most efficient at a QED vacuum resonance (VR) energy (typically around a few keV) that depends on the hydrogenic plasma density, the local field strength and light propagation direction, and acts to depolarize the net PD emerging from the surface.  Consequently, one expects a rotational-phase-averaged high PD away from the VR energy and significant PD reduction near it due to increased mode mixing.

\src\ is a canonical magnetar with a spin period of 2.09~s and a surface equatorial magnetic field strength of $\approx2.2\times10^{14}$~G \cite{camilo07ApJ:1547}.  It is currently in a persistently bright X-ray state, exhibiting a typical magnetar spectrum -- a soft thermal-like component prominent in the $\lesssim$10 keV band, and a hard X-ray tail dominating at higher energies \cite{cotizelati2020AA:1547,lower2023ApJ:1547}. The broadband 0.5--70 keV spectrum is best fit with either a double blackbody (BB) thermal contribution and a non-thermal power-law (PL) model or a single BB and double PL model \cite{cotizelati2020AA:1547,lower2023ApJ:1547}. The pulse shape at all energies is virtually sinusoidal, with minimal contribution from higher harmonics. The root-mean-squared (RMS) pulsed fraction \cite{bildsten97ApJ:PP} is notably large compared to the rest of the magnetar population -- it also exhibits a strong energy-dependence with a value of $\sim30$\% at 2~keV that steadily increases to 42\% around 6 keV, before decreasing monotonically towards 20\% at energies $>10$~keV  \cite{cotizelati2020AA:1547,lower2023ApJ:1547}. These spectral and timing characteristics likely indicate the presence of a thermal hot spot on the surface dominating the softest energies of the spectrum up to $\sim 6$~keV, whereafter emission of magnetospheric origin overtakes the spectrum's surface contribution and alters the pulse properties \cite{cotizelati2020AA:1547,lower2023ApJ:1547}. 

\src\ is also unique among the magnetar population as a persistent radio pulsar \cite{camilo2008ApJ}. It emits broad radio pulses at the spin-period of the magnetar, spanning approximately $25\%$ of the source duty cycle (Methods), unusually wide among the young, isolated NS population. The pulses are also highly polarized and display a characteristic S-shaped swing in the linear polarization position angle %(PA)
\cite{camilo2008ApJ}, indicative of the viewing line-of-sight crossing the radio emission cone \cite{zeng26ApJ:radiomag}. These properties hint at a nearly-aligned rotator, with an observer line-of-sight at a small angle to the rotation axis that samples open magnetic field lines proximate to the polar cap (Methods).

This richness of information offered by the combined X-ray + radio data motivated a deep simultaneous observing campaign with the Imaging X-ray Polarimetry Explorer (IXPE), the Neutron star Interior Composition ExploreR (NICER), and {\it Murriyang}, the 64 m Parkes radio telescope, which was carried out between March 26, 2025 UTC 03:01:52 and April 5, 2025 UTC 00:26:53. Our ultimate goal is to study in detail the soft (2--8~keV) X-ray polarimetric properties of the source as afforded by IXPE and to place it in context of the geometrical constraints achievable through the radio observations. Hence, all of the observations were coordinated to optimize simultaneous coverage between the instruments, particularly with IXPE (see Methods for full details of the observations).

The IXPE data reveal an energy and phase-averaged 2--8~keV PD of about $46\pm4\%$ ($1\sigma$ uncertainty; here and throughout the rest of the text) with a polarization angle (PA) of $-76\degree\pm 3\degree$ (Figure~\ref{fig:energy-resolved}) with a statistical significance of $\sim 10.5 \sigma$ above the minimum detectable polarization value for the 99\% confidence level (MDP$_{99}$). This is the largest such PD in the population of six IXPE-observed magnetars \cite{taverna24Galax, stewart2025ApJ:1841, rigoselli2025ApJ:1841}. The PD also displays strong energy dependence, achieving a maximum at $59\pm5\%$ in the softest energy band of 2--3 keV before decreasing to $37\pm5\%$ in the 3--4 keV range. The PA of the two bands, on the other hand, remain consistent with that of the energy-integrated measurement.  In the hard part of the energy range (4--8~keV), we measure a PD of $40\pm11\%$, and slightly shifted PA of $-86\degree \pm 8\degree$, a detection at only the $3.5\sigma$ significance level, mainly due to its lower signal-to-noise. Notably, the PD of the softest energy range is at least a factor of $2.5$ larger than the corresponding values measured in the rest of the population \cite{rigoselli2025ApJ:1841}.

Using a phase-coherent timing ephemeris that covers our full observing campaign, we phase-fold the X-ray (intensity and polarization) and radio data, after correcting for propagation through the ionized interstellar medium to ensure absolute alignment between the two wavelengths (see Methods). The PD is most prominent in the softer part of the spectrum (Figure~\ref{fig:phase-resolved}). It modulates moderately with phase, reaching a peak of $74\pm11\%$ in the 2--4 keV range and a minimum of about $34\pm8\%$; the former occurs close to the X-ray pulse minimum and away from the radio pulse, in contrast to the latter. Note that the PD reaches $82\pm15\%$ in the 2--3 keV range and a minimum of $42\pm12\%$: see Extended Figure~\ref{fig:3panel}. The PA displays a sinusoidal modulation between a maximum of $-51\degree\pm6\degree$ and a minimum of $-97\degree\pm9\degree$. Interestingly, the overall modulation is comparable in amplitude and shape to that observed in the radio band, albeit leading by a phase offset of around 0.3.  Notably, a similar phase offset is observed in the intensities, but now with the X-ray peak trailing the radio one.

A spectro-polarimetric analysis of the IXPE+NICER phase-averaged spectrum reveals that the emission in the soft 1--8~keV X-ray band is best described by a BB component with a temperature $kT=0.72\pm0.01$~keV and a relatively small emitting area $R^2=1.07 \pm 0.05$~km$^2$ (see Methods), assuming a source distance of 4.5 kpc \cite{tiengo10ApJ:1547}. In accentuating the model-independent polarization results presented above, we find that the phase-averaged BB polarization Q and U spectra are best modeled by a linearly decreasing PD from $65\pm8\%$ at 2 keV at a rate of $-15\pm2\%$~keV$^{-1}$, and a constant PA of $-67\degree ^{+4}_{-7}$. We do not find evidence for any trend of changing PA or PD at the upper-end of the IXPE energy band ($\gtrsim4$~keV), as hinted at (albeit at a low significance) in our model-independent analysis. This is likely due to the fact that the 2--8 keV IXPE source spectrum is heavily dominated by the 2--4~keV range, which contributes the most photons; indeed, adding a second component to the spectrum is not statistically justified, and, when included, remains unconstrained (see Methods).

The modest phase-averaged PD measured for most magnetars in the soft, 2--3~keV X-ray band \citep[$\lesssim 20\%$,][]{rigoselli2025ApJ:1841} has so far been interpreted as evidence for the presence of a condensed surface \cite{taverna22Sci, heyl2259}. The much higher PD we measure in \src\ challenges such a scenario for this source. Indeed, given the high surface temperature reported here and elsewhere \cite{cotizelati2020AA:1547, lower2023ApJ:1547} and a dipole field strength of $\approx2\times10^{14}$~G \cite{camilo07ApJ:1547}, magnetic condensation is unlikely for \src\ \cite{medin2007}. Phase-averaged polarization degrees approaching $\sim 40\%$ may be attainable in a pure magnetized atmosphere without invoking additional effects, but reproducing values of $\approx65\%$ is non-trivial \cite{taverna20MNRAS}. More challenging for such models are the extreme phase-resolved PDs ($\approx70$–$80\%$ in the 2--4 keV band; Figure~\ref{fig:phase-resolved}) and the absence of strong depolarization near the radio pulse phase. While we detect a minimum X-ray PD during the radio pulse, it remains comparatively high at $\approx30$–$40\%$, whereas substantial depolarization would be expected when sampling open field lines \cite{Dinh-2025-ApJ}. Together, these properties highlight tensions with standard surface and atmospheric models and suggest that additional physical processes influencing the propagation of polarized radiation through the magnetosphere may be at play.

The observed S-shaped sweep of the radio PA across the time and frequency averaged 2.5--4.0\,GHz profile (Figure~\ref{fig:phase-resolved}, top panel) is consistent with the rotating vector model (RVM) for a dipolar magnetic field geometry \cite{Radhakrishnan1969ApL}, and is well fit by this model, yielding a nearly aligned rotator viewed close to the spin axis (Extended Figure~\ref{fig:radioposteriors}); see Figure~\ref{fig:geometry} for a schematic of the geometry. We recover a magnetic and spin axes that are offset by $\alpha = 3.4\degree^{+1.3}_{-1.2}$, along with the angle between the spin axis and our line of sight of $\zeta  = 7.5\degree^{+3.0}_{-2.6}$. Under the simplifying assumption that the open field lines above the polar cap are fully illuminated, this combination of the magnetic and viewing geometry along with the high pulse duty cycle suggest the radio emission originates around $692^{+654}_{-400}$\,km above the NS surface (Methods). This height is less than $1\%$ of the light cylinder, indicating that the absolute radio pulse phase accurately describes a magnetic pole crossing.

The X-ray PA evolution can likewise be adequately described within the RVM formalism (see Methods and Figure~\ref{fig:xrayrvmpost}), indicating that the X-ray polarization follows a geometrical modulation consistent with the large-scale magnetic field structure. Although the constraints are weaker than in the radio due to the lower signal-to-noise, the inferred geometric parameters are broadly consistent with those obtained from the radio analysis at the $\sim2.5\sigma$ level, while not uniquely constraining the geometry \cite{li2026:pol22591547}. In particular, the allowed parameter space remains large, with only a limited region overlapping the radio-derived constraints \cite{li2026:pol22591547}. Differences between the radio and X-ray solutions, including phase offsets in the intensity and polarization profiles, suggest that the signals in the two bands originate from different locations relative to the magnetic axis of the star, and may arise from distinct emission altitudes or from deviations from a purely dipolar field configuration (see Methods).

The decrease in PD within the narrow energy range of 2--4 keV is notable. A separate, and differently polarized component contributing to the BB emission is not evident in the spectral and timing properties of the source which argue for the same thermal component dominating these energies. This behavior is most naturally attributed to mode conversion close to VR \cite{Lai-Ho-2003PhRvL, lai23PNAS}. Given the B-field strength of \src, and assuming a hydrogen atmosphere, mode conversion becomes efficient at energies $\gtrsim 2$~keV. At energies $\gtrsim 4$~keV, the interpretation becomes less robust due to the reduced signal-to-noise ratio. Nevertheless, the 4--8~keV (and 4--5~keV; see Methods) band shows a tentative deviation from the trend observed at lower energies. Deeper IXPE observations will be essential to clarifying the polarimetric behavior in this regime.

The exceptionally high polarization degrees reaching $65\pm8\%$ at 2 keV and rising to $82\pm15\%$ at certain rotational phases, the absence of strong depolarization near the radio pulse phase, the observed decrease in polarization in the 2–4 keV range, and the coherent, RVM-like evolution of the polarization angle across both X-ray and radio bands present a coherent observational picture. These elements, taken together, cannot be easily reconciled with the broad sampling of magnetic field directions represented in standard atmospheric models for extended surface X-ray hotspot locales. The signal properties are naturally interpreted in scenarios where the propagation of polarized radiation is influenced by quantum electrodynamic effects operating in the magnetosphere.  In particular, the combination of high polarization fractions and modest PA modulation suggests that the polarization direction remains closely coupled to the large-scale magnetic field geometry over extended distances. Within this context, magnetospheric VB provides a natural physical mechanism to account for the observed polarization behavior.

To illustrate the VB interpretation, we performed simulations of the X-ray intensity and polarization properties using the {\sl MAGTHOMSCATT} radiative transfer code (see Methods) \cite{Barchas-2021-MNRAS,Hu-2022-ApJ,Dinh-2026-ApJ}.  Within this adopted modeling framework, configurations that include magnetospheric VB provide a markedly improved description of the phase-resolved I, Q, U polarization data compared to models that neglect it (see Methods and Figure~\ref{fig:wedge-simulations}): non-refractive propagation yields notably stronger phase-dependent variations in Stokes $Q$ and $U$ than are present in the data.  Constrained by pulse phase offsets between the radio and X-ray signals (see Figure~\ref{fig:phase-resolved}), the best fit model indicates that the centroid of the X-ray hotspot is offset from the magnetic pole; see the diagram in Figure~\ref{fig:geometry}.  While these results are model-dependent, they provide additional support for the interpretation that magnetospheric propagation effects play a key role in shaping the observed polarization signatures in \src.

The combined X-ray and radio observational results presented here highlight the growing capability to probe QED effects in the extreme magnetic field environments of magnetars. Future X-ray polarimetric observations, particularly at lower energies where even higher polarization degrees may be expected, will be essential for further advancing this effort. Continued observations by IXPE, prospects afforded by future soft X-ray polarimetric missions such as GoSOX \cite{marshall21SPIE11822E}, together with laboratory and collider-based experiments, will enable increasingly stringent tests of the most fundamental predictions of quantum electrodynamics.

\section{Tables}
\begin{table}[h!]
	%\caption{\textbf{Spectro-polarimetric model}} 
	
	\centering
	\begin{tabular}{@{}cc@{}}
		\toprule
		\textbf{Model Parameter} & \textbf{Value $\pm$ $1\sigma$} \\ 
		\midrule
		$N_{\text{H}}$ (cm$^{-2}$)       & $3.67 \times 10^{22}$ (fixed) \\
		PD at 1 keV       & $0.80\pm 0.08$ \\
		PD$_{\text{slope}}$        & $0.15\pm0.02$ \\
		
		PA at 1 keV (deg)  & $-67^{+4}_{-7}$   \\
		PA$_{\text{slope}}$ (deg/keV) & $-4\pm 4$   \\
		BB kT (keV)            & $0.720\pm 0.005$ \\
		R$^2$ (km$^2$)                     & $1.07\pm 0.05$    \\
		%$\psi_{1}$ (deg)  & $-1\pm7$   \\
		%Flux             & $1.23\pm0.02$  \\
		
		C$_{\text{NICER  Obs 1}}$              & $1.0$ (fixed)  \\
		C$_{\text{NICER Obs 2}}$              & $1.08 \pm 0.05$  \\
		C$_{\text{NICER Obs 3}}$              & $1.05 \pm 0.05$  \\
		C$_{\text{NICER Obs 4}}$              & $1.01\pm 0.05$  \\
		C$_{\text{DU1}}$               & $0.80\pm 0.03$ \\
		C$_{\text{DU2}}$               & $0.80\pm 0.03$ \\
		C$_{\text{DU3}}$                & $0.76\pm 0.03$ \\
		Total Fit Statistic               & 1798.80 with 1511 d.o.f.\\
		%BB+PL Total Fit Statistic               & 2196.69 with 2138 d.o.f. \\
		\bottomrule
	\end{tabular}
	
	\label{tab:model_params}
	\caption{Spectral and polarimetric parameters for the best fit model, a single absorbed (column density, N\textsubscript{H}) blackbody (BB) with linearly varying polarization degree (PD) and angle (PA). A multiplicative cross-calibration uncertainty (labeled as C\textsubscript{instrument}) is also added to the model to track instrumental calibration difference across the four NICER observations and the three IXPE detector units (DUs). All uncertainties are calculated at the 68.3\% confidence level ($1\sigma$). The fit statistic employs a combination of the C-stat and Chi-Squared statistics used for NICER and IXPE, respectively. Blackbody emitting radius (R$^2$) is derived adopting a distance of 4.5 kpc \cite{tiengo10ApJ:1547}. } 
	
\end{table}

\newpage
\section{Figures}\label{figures}
\begin{figure}[H]
    \centering
    \includegraphics[width=0.6\linewidth]{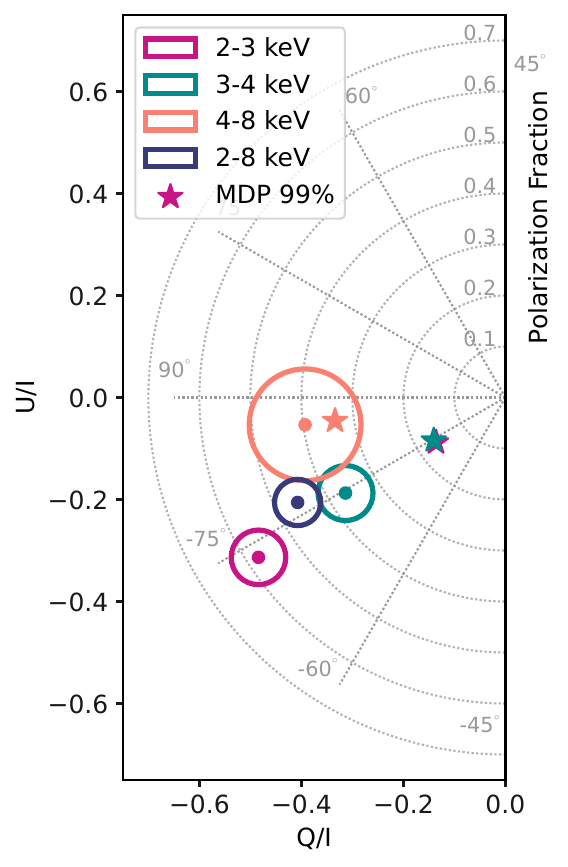}
    \caption{Background-subtracted and phase-averaged polarization measurements of \src. We display the PD in 4 energy bands, 2--3, 3--4, 4--8, and 2--8~keV. Contours show the $68.3\%$ ($1 \sigma$) confidence regions for Stokes Q/I and U/I (polarization degree and angle are shown in the radial and polar lines, respectively). The stars display the corresponding minimum detectable polarization at the 99\% confidence level (MDP$_{99}$) for each energy bin. The 2--8 keV energy-integrated polarization characteristics are shown in dark blue.  Notably, the energy-integrated PD (dark blue) is 46\% and the 2--3 keV PD (magenta) reaches 59\%. The PD decreases substantially in the energy range 3--4~keV. The PD remains high at 40\% in the 4--8 keV band, yet it suffers from a considerably lower signal-to-noise ratio (S/N). The PA does not exhibit strong variation across energy. }
    \label{fig:energy-resolved}
\end{figure}
\begin{figure}[H]
    \centering
    \includegraphics[width=\linewidth]{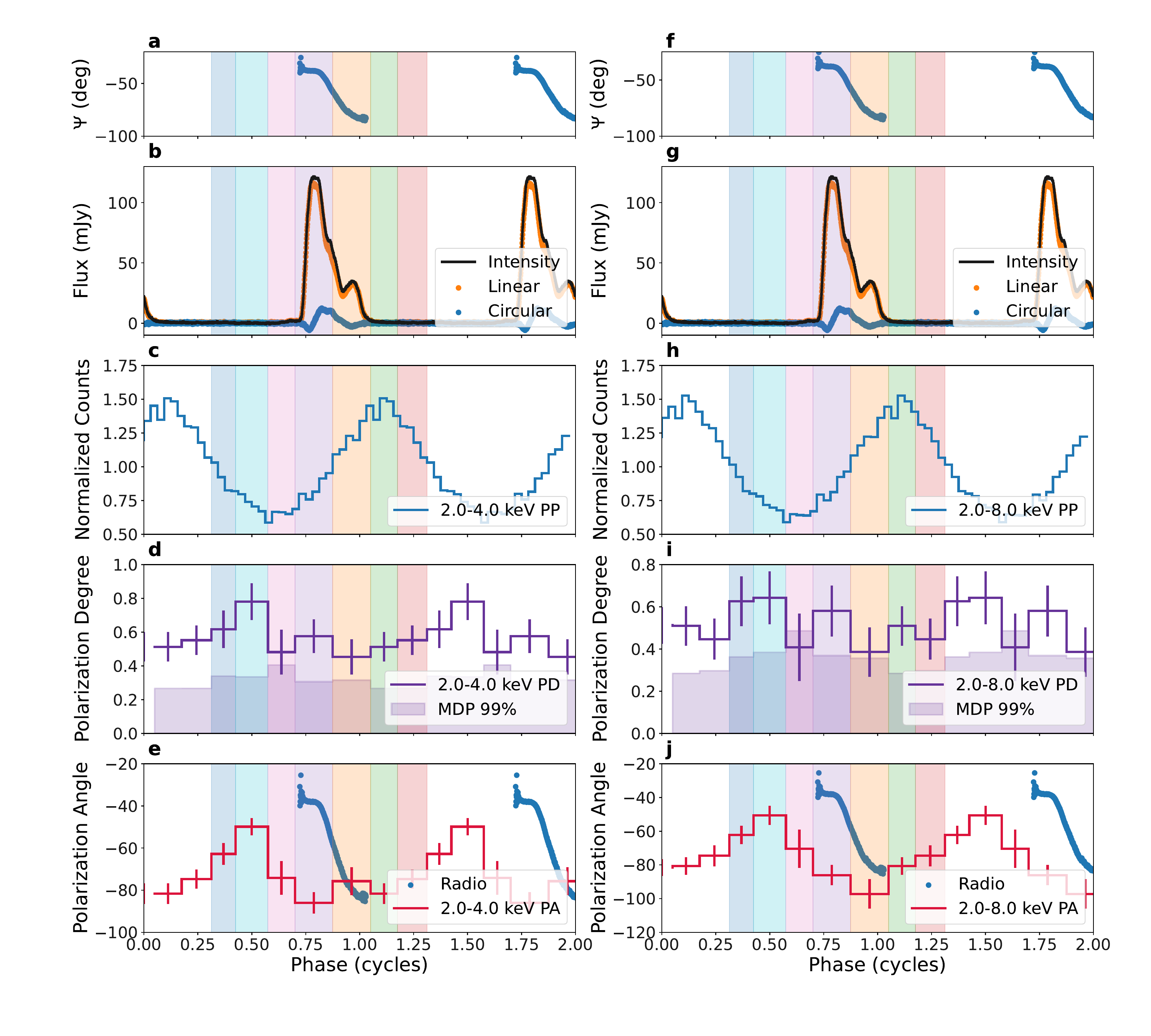}
   \caption{Phase-resolved radio and X-ray polarization characteristics of \src. The uppermost row shows the radio polarization angle (two cycles shown for clarity). The second row displays the radio intensity curve (black) along with the total linear radio polarization (orange) as well as the circular radio polarization (blue). All of the shown radio data are averaged across our three Murriyang observations. The bottom three rows show the IXPE intensity, polarization degree, and polarization angle pulse profiles, respectively. The radio PA is also displayed in the bottom panel for ease of comparison to the X-ray PA. The IXPE pulse profiles are displayed according to two energy bands: 2--4 keV (left column) and 2--8 keV (right column). The shaded curves overlaying the polarization degree panels depict the MDP$_\text{99}$ at each corresponding phase bin. The vertical colored bands highlight the binning used for the IXPE phase-resolved polarization analysis.}
    \label{fig:phase-resolved}
\end{figure}

\begin{figure}[H]
	\centering
	\includegraphics[width=0.9\linewidth]{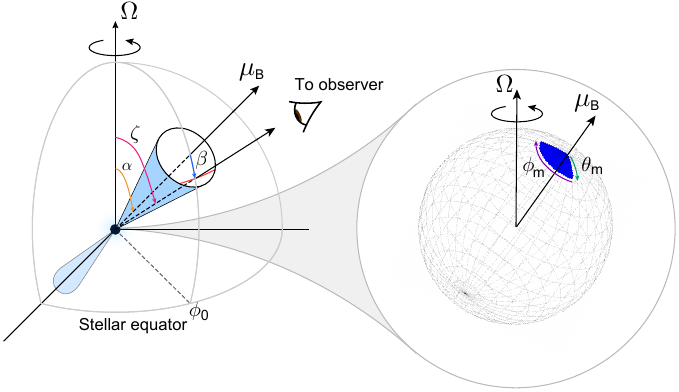}
		\caption{{Illustration of a rotating NS with a dipole magnetic field geometry described by the rotating vector model.} The magnetic axis ${\mu}_{\rm B}$ inclined from spin axis $\Omega$ by an angle $\alpha$ and an observer viewing angle from the spin axis $\zeta$. The magnetic meridian is defined at a reference phase $\phi_{0}$ (on left), for which the spin and magnetic axes and the observer line of sight (l.o.s.) are coplanar.   The emission radio cone (medium blue) is centered around ${\mu}_{\rm B}$ and its intersection with the observer l.o.s. is indicated by the red line. Our best-fit wedge-shaped hot spot corresponding to the light curves in Figure~\ref{fig:wedge-simulations} is depicted in blue in the inset on the right. The hotspot extent on the NS surface in the inset is constrained by a defined magnetic co-latitudes $0 \leq \theta_m \leq 17\degree$ and longitudes $\phi_m$ ranging from the meridian to 120$\degree$. }
	\label{fig:geometry}
\end{figure}

\section{Methods}\label{sec:methods}
%TC:endignore
\subsection{X-ray Spectro-polarimetric Analysis}

We performed spectro-polarimetric analysis of the 0.5--12 keV NICER and 2--8 keV IXPE spectral data using Xspec version 12.15.0 to investigate the nature of the source emission. For the NICER (IXPE) spectra, we grouped each spectral bin with a minimum of 5 (30) counts per bin and accordingly performed statistical fits with the W-statistic (Chi-squared) via \texttt{c-stat} (\texttt{chi}) within Xspec. While the energy channels typically contain substantially more counts than these minimum thresholds, the adopted grouping ensures that the lowest and highest energy bins remain in the appropriate statistical regime for each instrument. Given the source's low S/N at energies $\lesssim 2$ keV, we froze the column density (N$_\text{H}$) to a value of $3.67\times 10^{22}$ cm$^{-2}$ as reported by Ref. \cite{lower2023ApJ:1547}; we adopted the Wilm solar abundances \cite{wilms00ApJ} and Verner cross-sections \cite{verner96ApJ:crossSect} to preserve consistency with the previous analysis. Finally, our spectral modeling imposes a constant between each telescope to track cross-instrument calibration uncertainties (see Table \ref{tab:model_params}).

We simultaneously fit the NICER SCORPEON background model with the source spectra. While fitting, we fix the trapped-electron component normalizations to 0 since they negligibly contribute to the fitted background. The background models reveal that the NICER source spectra is background-dominated below $\approx 1.4$ keV and above $\approx 7$ keV. Likewise, the background dominates the IXPE spectra above $\approx 6$ keV, limiting the statistics of the higher energy contributions. 

Historically, the broadband spectrum has been well described using either a 2BB+PL or a BB+2PL fit \cite{cotizelati2020AA:1547, lower2023ApJ:1547}. However, the source data’s limited statistics constrain the detectability of the hot BB or PL components which emerge at $\gtrsim 6$ keV. We performed an F-test for a single and double BB model and found an F-statistic value of 2.47 and associated probability of 0.09. While the F-test favors including the second BB component, the 2BB best fit model yields physically unrealistic values. We thus reject the extra component in favor of a single absorbed BB model for this particular data set, with a total-fit statistic of 1798.80/1511 d.o.f (see Extended Figure \ref{fig:spectra}-\ref{fig:nicer_spectra}). Additionally, we favor applying a linear polarization component to the model with a total fit statistic of 1798.80/1511 d.o.f. compared to 1946.38 for a power-law component and 1840.09 for a constant component for 1513 d.o.f. A single absorbed power-law model yielded largely unphysical parameters and thus was rejected.

A significant detection was found in the 4--8 keV band of the model-independent analysis (Figure \ref{fig:energy-resolved}). However, the high-energy component possesses large uncertainties, impeding comparisons to the 2--3 and 3--4 keV behavior and in discerning trends in the energy-resolved polarization characteristics. We further investigated the high-energy component by subdividing the energy bands to 4--5 keV and 5--8 keV, similar to the analysis presented in \cite{Taverna_2026}. We analyzed these bins comparatively using a model-independent approach in \texttt{ixpeobssim} and a model-dependent approach in Xspec wherein each energy bin was fit according to an absorbed BB model with a constant polarization component. We fixed the BB model parameters to their best-fit value obtained from the energy-integrated NICER+IXPE spectral fits. This bin-by-bin spectro-polarimetric approach is standard practice in IXPE analyses and effectively yields model-independent polarization measurements. The Xspec error contours were obtained by Monte Carlo sampling of the best-fit parameter confidence levels and transforming the resulting distributions into Stokes Q-U space. The \texttt{ixpeobssim} and Xspec analyses yield a PD of approximately $50\pm10\%$ and $48\pm 15\%$ respectively, and the \texttt{ixpeobssim} analysis reveals that the 4--5 keV band exhibits a statistically significant detection at the $\sim5.5\sigma$ level (see Extended Data Figure \ref{fig:Stokes_Q_U_polar}). While the 4--5 keV values appear to deviate from the lower-energy trend, their large uncertainties remain consistent with the 2--3 and 3--4 keV bands. The  2--4 keV band dominates the observed counts while the 4--8 keV band contributes $\sim10\%$, possibly explaining why the linear polarization model fits the 2--8 keV data well.  A deeper exposure is required to constrain the data above 5 keV, where there may likely be a contribution from another spectral component \cite{cotizelati2020AA:1547, lower2023ApJ:1547}. 

We use the Goodman-Weare MCMC method within Xspec with 10 walkers drawn from a multidimensional gaussian centered on the best-fit parameters. Three chains of 50000 simulated spectral fits were generated (totaling at 150000) with the first 5000 steps from each chain burned. The resulting posterior distributions were used to generate 1- and 2-D corner plots and assess parameter correlations, shown in Extended Figure~\ref{fig:mcmc}. We used Xspec’s standard uniform priors for each model parameter, which sufficiently sampled the full physically-motivated parameter space apart from the PA slope where we expanded the prior limits from $\pm5\degree$/keV to $\pm30\degree$/keV.

\subsection{Radio Polarization Analysis} \label{subsec:radio}

Assuming the radio emission from \src\ originates from above the polar cap of a dipole magnetic field component \cite[e.g., ][]{zeng26ApJ:radiomag}, we can use the rotating vector model to constrain the magnetic and viewing geometry of the magnetar \cite{Radhakrishnan1969ApL}. The linear PA sweep can be described as
\begin{equation}
    \tan(\psi - \psi_{0}) = \frac{\sin\alpha \sin(\phi - \phi_{0})}{\sin\zeta\cos\alpha - \cos\zeta\sin\alpha\cos(\phi - \phi_{0})},
 \label{eq:RVM_angles}
\end{equation}
where $\psi_{0}$ is the PA at the inflection point of the RVM swing, $\phi$ is the rotation phase of the magnetar, $\phi_{0}$ is the phase at which the PA inflection occurs, $\alpha$ is the magnetic inclination angle and $\zeta = \alpha + \beta$ is the viewing angle between the spin-axis and our line of sight, which is the product of the magnetic inclination angle and impact parameter ($\beta$). Note that the PA swing of the RVM is inverted, i.e $\psi_{\rm obs} = -\psi_{\rm RVM}$ in the observer reference frame \cite{Everett2001ApJ}. We fit the RVM to the average profile obtained from combining our three concurrent \textit{Murriyang} observations using the {\tt bilby} Bayesian inference library \cite{2019ApJS..241...27A} as an interface to the {\tt dynesty} nested sampling algorithm \cite{2020MNRAS.493.3132S}, and a Gaussian likelihood of the form
\begin{equation}
    \mathcal{L}(d | \theta) = \prod_{i}^{n} \frac{1}{\sqrt{2\pi\sigma_{i}^{2}}} \exp \Big[ -\frac{(d_{i} - \mu(\theta))^{2}}{2\sigma_{i}^{2}} \Big],
\end{equation}
where $d$ is the data (corresponding to the PA), $\mu(\theta)$ is the RVM with model parameters given by $\theta$, and $\sigma$ is the uncertainty on the PA. Note, we modified the formal uncertainties on the PA by adding in an additional error in quadrature term (EQUAD) as $\sigma^{2} = \sigma_{\psi}^{2} + \sigma_{\rm EQUAD}$ to account for potential underestimation of the PA uncertainties due to uncorrected systematic effects. We assumed priors on $\alpha$ and $\zeta$ that were uniform in solid angle, i.e where ($\pi(\theta) \propto \sin\theta$). Standard uniform priors were used for $\phi_{0}$ and $\psi_{0}$. Our prior bounds for each parameter were set between $0\degree < \pi(\alpha) < 180\degree$, $0\degree < \pi(\zeta) < 180\degree$, $270\degree < \pi(\phi_{0}) < 360\degree$ and $-90\degree < \pi(\psi_{0}) < 0\degree$.
Note, the restricted prior range on $\phi_{0}$ is required to break a symmetry in the RVM that results from the unknown rotational direction (prograde or retrograde) of the magnetar. The resulting one- and two-dimensional posterior distributions are displayed in Extended Figure~\ref{fig:radioposteriors}. 

Our recovered values of { $\alpha = 3.4\degree^{+1.3}_{-1.2}$} and {${\zeta = 7.5\degree^{+3.0}_{-2.6}}$} (equating to {${\beta = 4.2\degree^{+1.7}_{-1.5}}$}) indicate 1E~1547.0$-$5408 is an aligned rotator, viewed almost pole-on. This geometry is seemingly in tension with the previously reported values of $\alpha = 160\degree$ and $\beta = 14\degree$ \cite{camilo2008ApJ}. However, it can be largely explained by the covariance between the magnetic and viewing angles due to the unknown spin direction of the magnetar, and a potential usage of the incorrect PA sign convention in their RVM fit. Re-fitting the PA swing of the 2007 observation analyzed by Ref. \cite{camilo2008ApJ}, we obtained values of {${\alpha = 10.2\degree^{+3.9}_{-3.5}}$} and {${\zeta = 17.7\degree^{+6.7}_{-6.1}}$}, for a magnetic impact angle of {${\beta = 7.5\degree^{+2.9}_{-2.6}}$}. This updated fit is consistent with our 2025 measurement to within the 68--95\% confidence intervals. When the incorrect PA sign convention is used, we obtain similarly large values of ${\alpha = 169.7\degree^{+3.5}_{-4.0}}$ and {${\zeta = 162.2\degree^{+6.1}_{-7.0}}$} to those of Ref. \cite{camilo2008ApJ}.

An estimate of the radio emission height can be inferred from a combination of the profile phase width ($W$) and the measured magnetic/viewing geometry.  Assuming a filled emission cone, the beam-opening angle ($\rho$) can be computed as \cite{1984A&A...132..312G}
\begin{equation}
    \cos\rho = \cos\alpha\cos\zeta + \sin\alpha\sin\zeta\cos(W/2).
\end{equation}
The opening angle is related to the emission height ($h_{\rm em}$) via \cite{1990ApJ...352..247R}
\begin{equation}
    \rho = 3\sqrt{\frac{\pi\,h_{\rm em}}{2Pc}},
\end{equation}
where $P$ is the magnetar rotation period and $c$ is the vacuum speed of light.
Typically the value of $W$ is taken to be either the profile width at the 10\% or 50\% (i.e $W_{10}$ or $W_{50}$) levels.
However, given the breadth of the pulse peak of \src\ (Figure~\ref{fig:phase-resolved}), after iterating over different on-pulse window sizes, we settled on the 1.5\% profile width of $W = W_{1.5} = 142\degree$. Combining this value with our measurements of $\alpha$ and $\zeta$, we obtain {${h_{\rm em} = 692^{+654}_{-400}}$\,km}. 

This is broadly consistent with emission heights inferred for rotation-powered pulsars \cite{2023MNRAS.520.4801J}, yet substantially lower than those deduced for other magnetars \cite{2021MNRAS.502..127L, 2024NatAs...8..617D}.

\subsection{X-ray Polarization Angle Modeling} \label{subsec:x-ray-rvm}

We model the X-ray PA swing using the RVM following the same formalism adopted for the radio analysis (Equation~\ref{eq:RVM_angles}). The data are restricted to the 2–4 keV energy band, where the signal-to-noise ratio is highest (see Figure~\ref{fig:energy-resolved}).

We performed a Bayesian inference of the RVM parameters with the ensemble sampler {\tt emcee} \cite{Foreman13PASP:emcee}. Following previous IXPE analyses of X-ray pulsars \citep[e.g.,][]{doroshenko22NatAs}, we used the wrapped-angle likelihood of \cite{NKC1993AA:xrayrvm}. This likelihood is specifically designed to handle the non-Gaussian uncertainties inherent to PA measurements and remains statistically valid in the presence of low–signal-to-noise data points. The phase-resolved PA profile is binned into 15 rotational bins. We used the same priors as the radio RVM fits, albeit with extended prior ranges on $\psi_{0}$ and $\phi_{0}$ of $-180\degree < \pi(\psi_{0}) < 0\degree$ and $0\degree < \pi (\phi_0) < 360\degree$.

The RVM provides an adequate description of the data, with $\chi^2=12.8$ for 11 d.o.f. ($\chi^2_\nu=1.16$). The resulting posterior distributions are shown in Extended Figure~\ref{fig:xrayrvmpost}. Given the lower photon statistics compared to the radio data, the X-ray constraints are broader, particularly for the viewing angle $\zeta=69\degree_{-23\degree}^{+25\degree}$ (Extended Figure~\ref{fig:xrayrvmpost}). The inferred geometric parameters are broadly consistent with those obtained from the radio RVM fit at the $\sim2.5\sigma$ level; however, the X-ray data do not uniquely constrain the geometry. In particular, the allowed parameter space remains large, with only a limited region overlapping the radio-derived constraints. Within these uncertainties, we recover a small magnetic inclination angle, $\alpha=18\degree\pm4\degree$ ($68\%$ confidence interval), consistent with a quasi-aligned geometry. To assess the robustness of this result, we repeated the analysis using phase bins ranging from 8 to 16, obtaining consistent posterior distributions in all cases. We further validated our findings using an unbinned likelihood formalism \citep{gonzalez23MNRAS:rvm}, which yielded fully consistent parameter estimates. We note that recent Bayesian RVM modeling of the X-ray PA swing in \src, utilizing priors that are informed by the radio best-fit parameters, returns posteriors that are fully compatible with the radio-inferred geometry, that of a quasi-aligned rotator \cite{li2026:pol22591547}.

We find a large and statistically significant difference between the X-ray and radio fit for $\phi_0$ amounting to $\Delta\phi_0\approx80\degree$. This may be attributed to several contributing factors. For uniform cone emission, the radio $\phi_0$ marks the rotational phase at which the magnetic axis is closest to the line of sight, which arises at the meridional plane where the spin axis, magnetic axis and observer's line of sight are coplanar.  In contrast, the X-ray surface spot for, e.g., the wedge case, need not be uniform in magnetic longitude and with a centroid that is offset from the pole.  Such an angular displacement between the X-ray and radio emission locales is forced by the phase lag in the respective intensity peaks. The $\Delta\phi_0$ value likely receives an additional contribution from the distinct altitudes at which the polarization vectors are realized in the two wavebands. Our radio analysis suggests emission heights of order 50–100 stellar radii, whereas the polarization-limiting radius inferred for the X-ray emission likely lies at only a few tens of stellar radii (see Section~\ref{sec:xray_modeling} below). This could be further impacted by any toroidal twists to the field morphology at the respective altitudes.

The reference position angle $\psi_0$ inferred from the X-ray RVM fit differs from the radio value by $\approx15$\degree, corresponding to a fractional difference of $\lesssim$20\%. For a symmetric radio cone, the parameter $\psi_0$ corresponds to the projected orientation of the spin axis on the plane of the sky, modulo an additive constant associated with the dominant polarization eigenmode. For an ideal RVM in a pure dipole geometry, similar emission heights, and no propagation effects, the radio and X-ray reference PAs should either be identical if the two emissions are dominated by the same polarization mode or offset by 90\degree\ if they are not. The observed difference is modest enough to suggest that the radio and X rays are indeed dominated by the same polarization mode.  It is likely that its non-zero value could be the imprint of magnetospheric field twists. The principal signature of twists is that they will rotate the PA (i.e., angle $\psi_0$) due to the presence of a toroidal field component. Given somewhat distinct emission heights for the two wavebands, modest variance in PA is expected \cite{hibschman2001ApJ, tong21MNRAS:rvm}.

In conclusion, the ability for the RVM to adequately describe the X-ray PA and the radio one provides an important consistency check for the VB interpretation. In the presence of strong QED birefringence, the X-ray polarization vector is expected to adiabatically track the local magnetic field direction out to the polarization-limiting radius, beyond which it becomes fixed. This naturally produces an RVM-like PA swing. The observed similarity between the X-ray and radio PA morphologies supports a scenario in which the large-scale dipolar field geometry, perhaps combined with a moderate contribution from twists, governs the polarization in both bands, with VB acting to preserve this imprint in the X-ray signal.

\subsection{Interpreting the X-ray Intensity and Polarization Signals} \label{sec:xray_modeling}

To model the anisotropy and polarization characteristics of X-ray emission observed for \src, we employed {\sl MAGTHOMSCATT} \cite{Barchas-2021-MNRAS,Hu-2022-ApJ,Dinh-2025-ApJ,Dinh-2026-ApJ}, a Monte Carlo radiative transfer simulation that treats polarized scattering transport in magnetized NS atmospheres.  At any point on the surface removed from the magnetic pole, the emergent radiation is highly polarized ($\gtrsim 80\%$); see Ref.~\cite{Dinh-2025-ApJ} for details on the angular distributions of local anisotropy and polarization.  General relativistic (GR) light bending and redshifting are included: herein we adopted a compactness value of 0.425 for a NS radius of $R_{\rm NS} = 12$~km and mass of $M_{\rm NS}\approx1.7 $M$_{\odot}$. The influences of VB during magnetospheric propagation are encoded and can be switched on or off, as desired \cite{Dinh-2026-ApJ}.  Mode conversion at the VR deeper in the atmosphere is not incorporated; we only model the softest energy band of $2-3$~keV, where the impact of VR is expected to be relatively insignificant.

The simulation is versatile in treating local atmospheric magnetic field directions in GR dipole geometry, allowing for modeling various hot-spot configurations. We explored various values of the hot-spot area $A$ and found that the conclusion about VB was not dependent on our specific choice of $A$. Thus, we restrict the discussion here to  $A = A_{\rm BB}$  cases only, where $A_{\rm BB} = 13$~km$^2$ is the inferred BB emission area from the best spectral fit.  Different hot-spot configurations, both circular and non-circular, were considered. For circular hot spots, an area of $A_{\text{BB}}$ corresponds to a half-opening angle of $\theta_{\rm cap} \approx  10\degree$. For non-circular ones, we assume a wedge for simplicity. The inclination angle $\alpha$ ranged from $0\degree$ (aligned rotator) to $90\degree$ (orthogonal rotator) in increments of $2\degree$. The viewing angle $\zeta$ ranged from $0\degree$ to $180\degree$, divided into 180 bins.

Among all cases considered, the best fit to the intensity {(${\chi^2/\rm{dof} = 33.9/30 }$)}, Stokes $Q/I$ {(${\chi^2/\rm{dof} = 19.0/4 }$)}, and $U/I$ {(${\chi^2/\rm{dof} =10.9/4} $)} data correspond to an offset non-circular configuration, represented by a wedge with magnetic co-latitudes spanning $0 \leq \theta_m \leq 17\degree$ and longitudes ranging from the meridian to $120\degree$ (see the blue region in Figure~\ref{fig:geometry}).  This was obtained for $(\alpha, \zeta) = (0\degree, 1.5\degree)$ with the inclusion of magnetospheric VB; see the blue curves in Extended Figure~\ref{fig:wedge-simulations}. We note that the dominant contribution to the $\chi^2$ values in the $Q/I$ and $U/I$ fits comes from the 7th phase bin  ($\phi=0.875$), which is only slightly above the $99\%$ MDP. The $(\alpha, \zeta)$ values  inferred through our modeling approximately support the aligned geometry inferred by the radio intensity and the radio RVM, yet are discrepant with the X-ray RVM best-fit results. A detailed discussion of these differences is beyond the scope of this work; they may reflect a combination of non-dipolar field structure, and distinct emission altitudes for the X-ray and radio signals. 

The $Q/I$ and $U/I$ traces without VB (Extended Figure~\ref{fig:wedge-simulations}, orange curves) show strong variations when compared to the VB-on case, a consequence of the convolution of disparate magnetic field directions on the stellar surface in conjunction with non-refractive propagation through the magnetosphere. This VB-off scenario does not provide an adequate fit to the data (panel b, ${\chi^2/\rm{dof} = 106.8/4 }$, and panel c, ${\chi^2/\rm{dof} = 95.6/4 }$). We also show in panel (d) the best case among the VB-off models corresponding to a pole-centered circular hot spot (red). This too does not adequately reproduce the observed polarization data, and the corresponding geometry $(\alpha, \zeta = 2\degree, 20.5\degree) $ is inconsistent with that inferred in the radio. Hence, within our adopted atmospheric modeling framework, the inclusion of magnetospheric VB provides a better description of the X-ray intensity and polarization measurements in \src.

This inference of the likely action of VB in the magnetosphere can be made regardless of which polarization eigenmode (O-mode, as in {\sl MAGTHOMSCATT}, or X-mode, as in \cite{Lai-Ho-2003PhRvL,vanadelsberg09MNRAS, 2002ApJ...566..373L, 2006MNRAS.373.1495V}) emerges in dominance from the atmosphere.  Using the wedge configuration (Extended Figure~\ref{fig:wedge-simulations}), we tested this by performing simulations where the polarization degree was presumed uniform across the surface and $\Pi_X = I_X/(I_X+I_O)$ was allowed to float as a free parameter on the interval $[0,1]$. We found that the fits to the $Q/I$ and $U/I$ data were poor when $0.2 \lesssim \Pi_X \lesssim 0.8$, for both VB on and off, while fits for $\Pi_X < 0.2$ (O-mode dominance) and for $\Pi_X > 0.8$ (X-mode dominance) worked well only with VB on (and equally so), and poorly with VB off.  This exploration will be detailed in a future study.

\renewcommand\tablename{Extended Data Table}
\setcounter{figure}{0} 
\setcounter{table}{0} 
\renewcommand\figurename{Extended Data Figure}

\section{Extended Data}
\begin{figure}[H]
	\centering
	\includegraphics[width=0.9\linewidth]{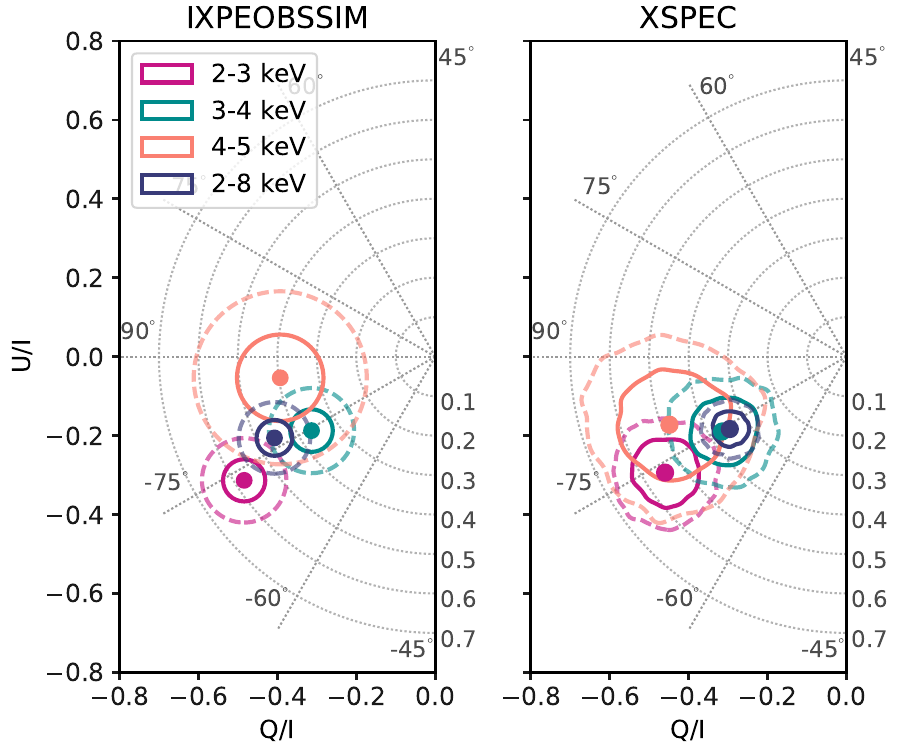}
	\caption{A comparative study between model-independent Stokes Q and U polarization characteristics obtained through using IXPEOBSSIM (left) and through applying a \texttt{polconst*bbodyrad} spectral model in Xspec (right) at four energy bins: 2--3 keV, 3--4 keV, 4--5 keV, and 2--8 keV. The resulting PD and PA for the two methods are comparable to each other. Notably, both show a non-linear trend in the energy-dependence of the PD as the 4--5 keV band approaches the value of the 2--3 keV band at $\sim1\sigma$ level. Future observations with higher count statistics are needed to complement this study to more directly probe the nature of the high-energy polarization. }
	\label{fig:Stokes_Q_U_polar}
\end{figure}
\begin{figure}[H]
	\centering
	\includegraphics[width=0.9\linewidth]{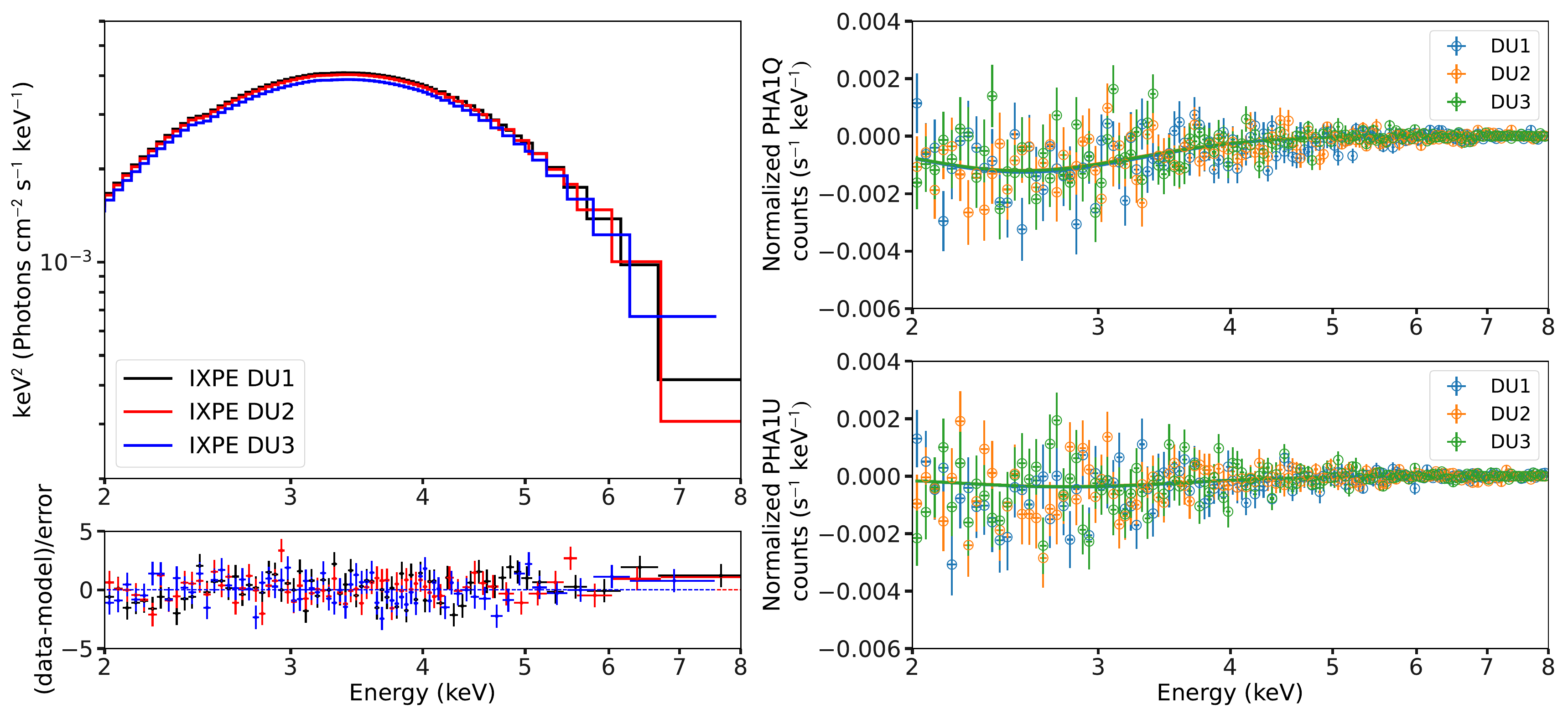}
	\caption{The simultaneous spectral $\nu F_{\nu}$ models of the NICER+IXPE\ observations. The upper left panel displays the best fit model, a single absorbed blackbody with a linear polarization component: \texttt{constant*tbabs(pollin*bbodyrad)}. NB: Only the IXPE spectra are displayed here for the sake of visual clarity. The bottom left panel shows the data divided by the folded model for the absorbed BB model. The right-hand panels show the normalized Stokes \textit{Q/I} (top) and \textit{U/I} (bottom) spectra in linear space for the three IXPE DUs with the solid lines showing the best fit of the linear polarization component. The quasi-thermal BB is accompanied by a strong polarization signal that decreases as a function of energy.  }
	\label{fig:spectra}
\end{figure}
\begin{figure}[H]
	\centering
	\includegraphics[width=0.65\linewidth]{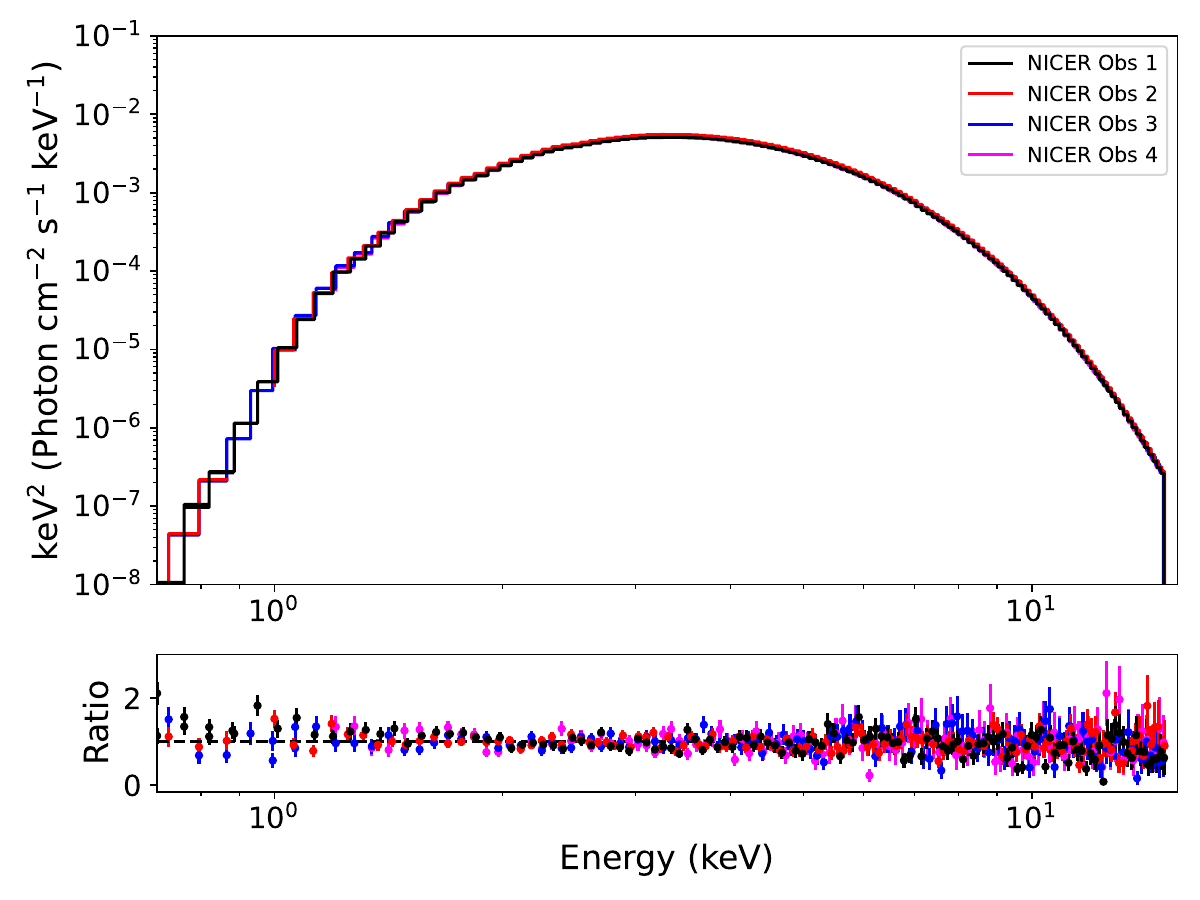}
	\caption{Simultaneous spectral modeling of the NICER+IXPE observations in the $\nu F_{\nu}$ representation (see also Figure~\ref{fig:spectra}). The top panel shows the best-fit model to the NICER spectra, described by a single absorbed blackbody. For visual clarity, we do not display the SCORPEON background model components. The bottom panel shows the ratio of the NICER data to the folded model.}
	\label{fig:nicer_spectra}
\end{figure}
\begin{figure}[H]
	\centering
	\includegraphics[width=.9\linewidth]{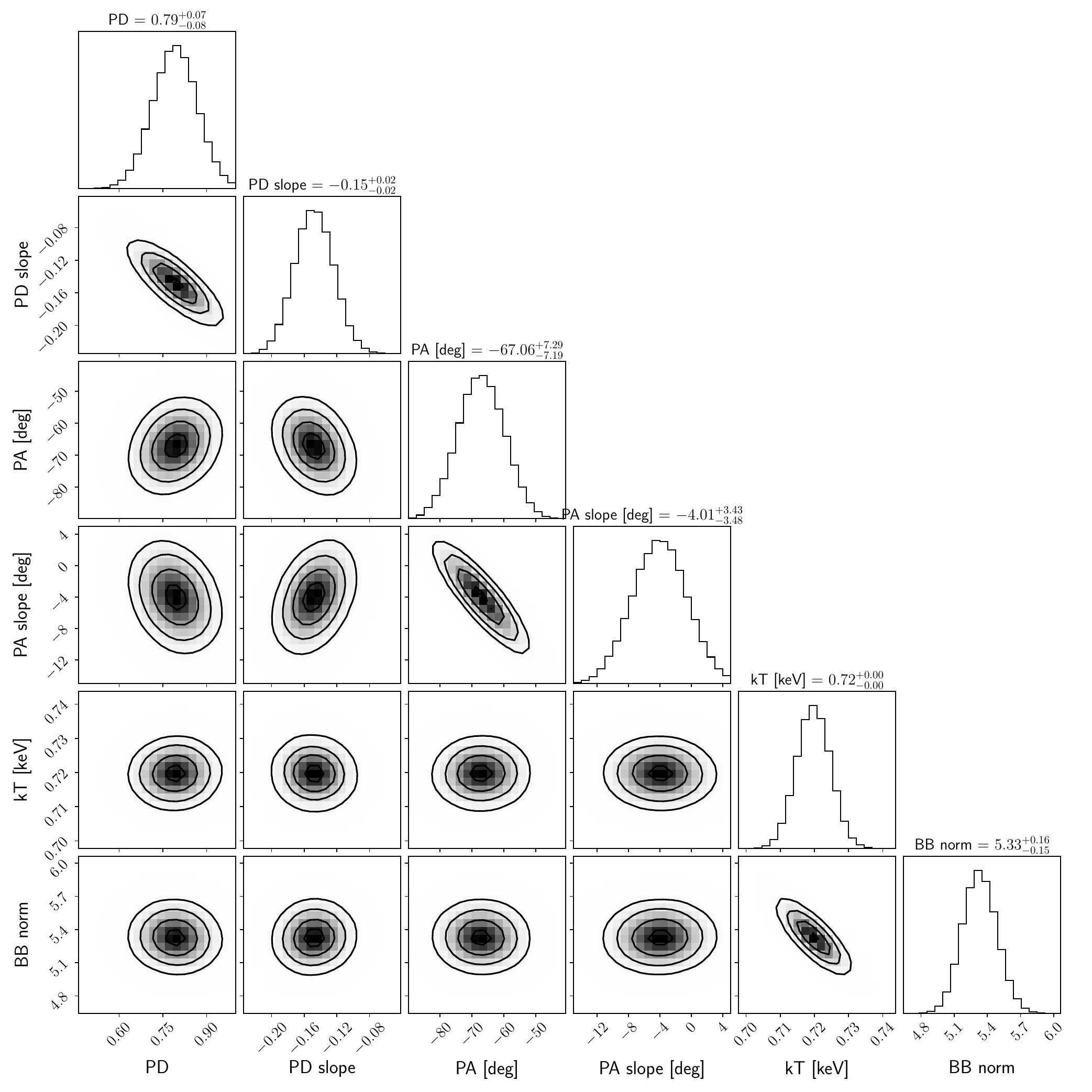}
	\caption{Corner plots of the one- and two-dimensional posterior distributions of the X-ray spectro-polarimetric parameters produced by an MCMC chain. The contours denote the 11.8\%, 39.3\%, 67.5\%, 86.4\% credible regions (corresponding to 0.5, 1, 1.5, and 2$\sigma$).}
	\label{fig:mcmc}
\end{figure}
\begin{figure}[H]
	\centering
    \includegraphics[width=.9\linewidth]{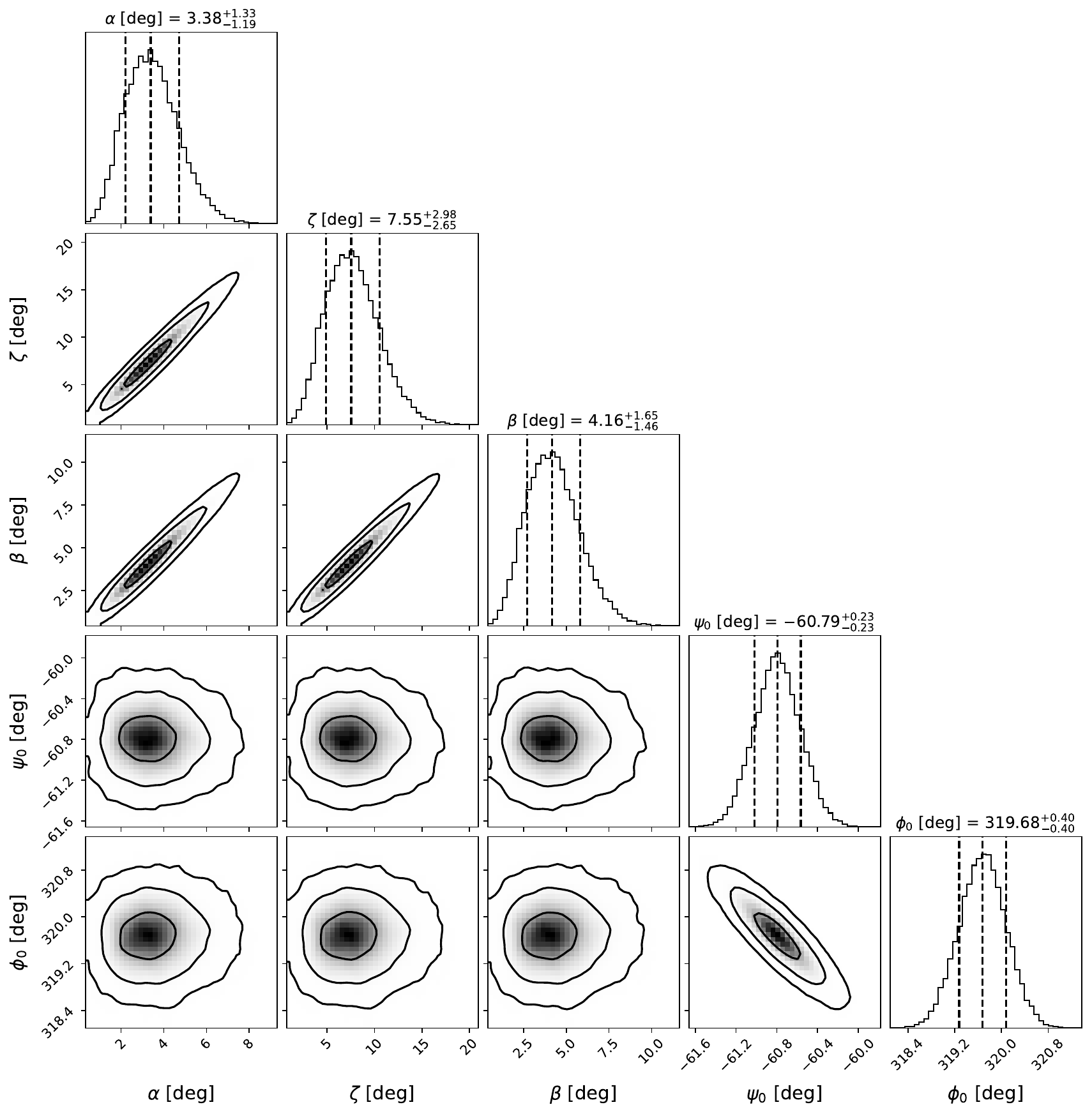}
\caption{Corner plots of the one- and two-dimensional posterior distributions of the radio RVM fit parameters. The contours denote the 39\%, 87\%, and 99\% credible regions (corresponding to 1, 2, and 3$\sigma$).}
	\label{fig:radioposteriors}
\end{figure}
\begin{figure}[H]
	\centering
	\includegraphics[width=\linewidth]{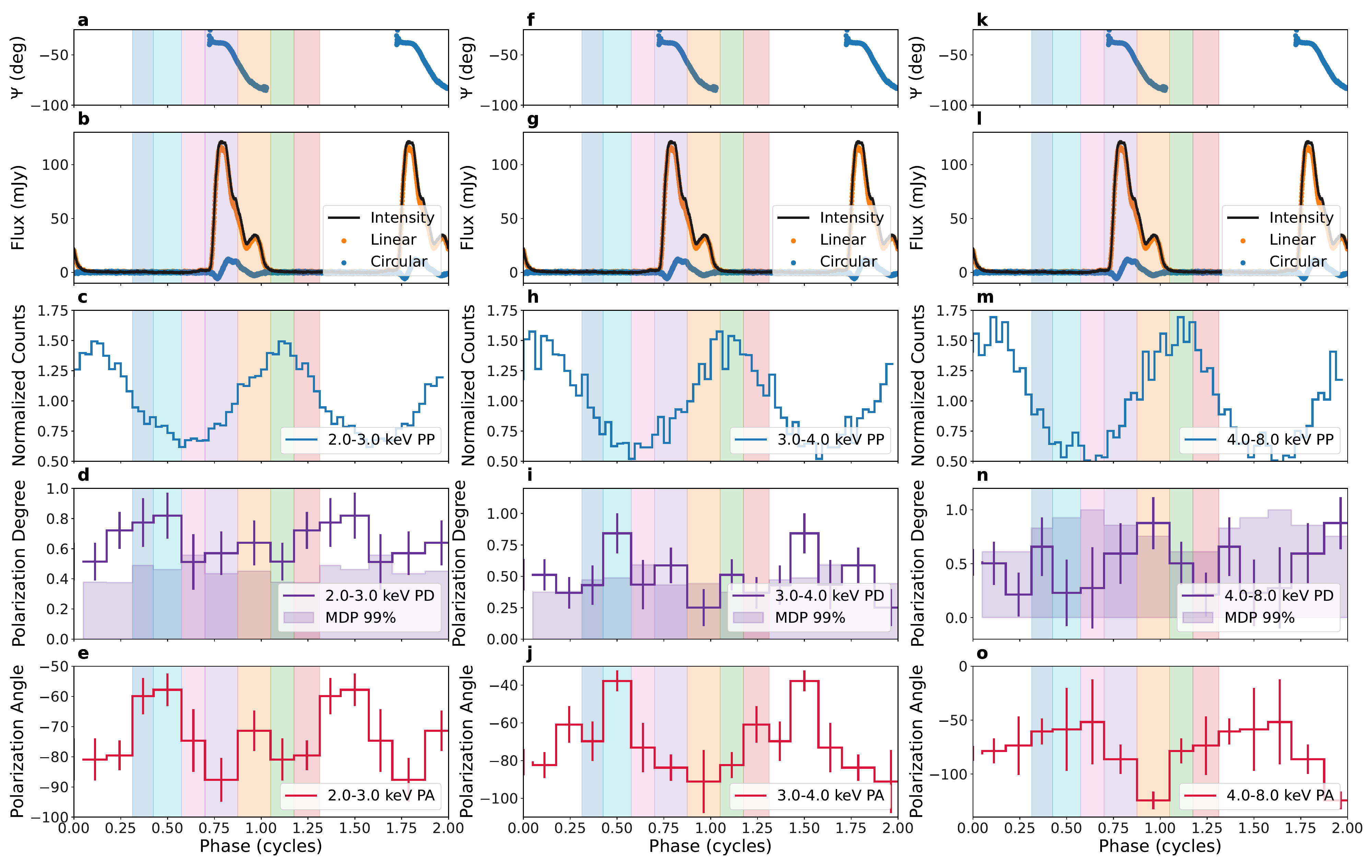}
 \caption{Complementary plot to Figure~\ref{fig:phase-resolved}, with IXPE intensity, PD, and PA (shown in the lower three panels, respectively) binned according to 2--3 keV (left), 3--4 keV (center), and 4--8 keV (right).}
	\label{fig:3panel}
\end{figure}

\begin{figure}[H]
	\centering
\includegraphics[width=\linewidth]{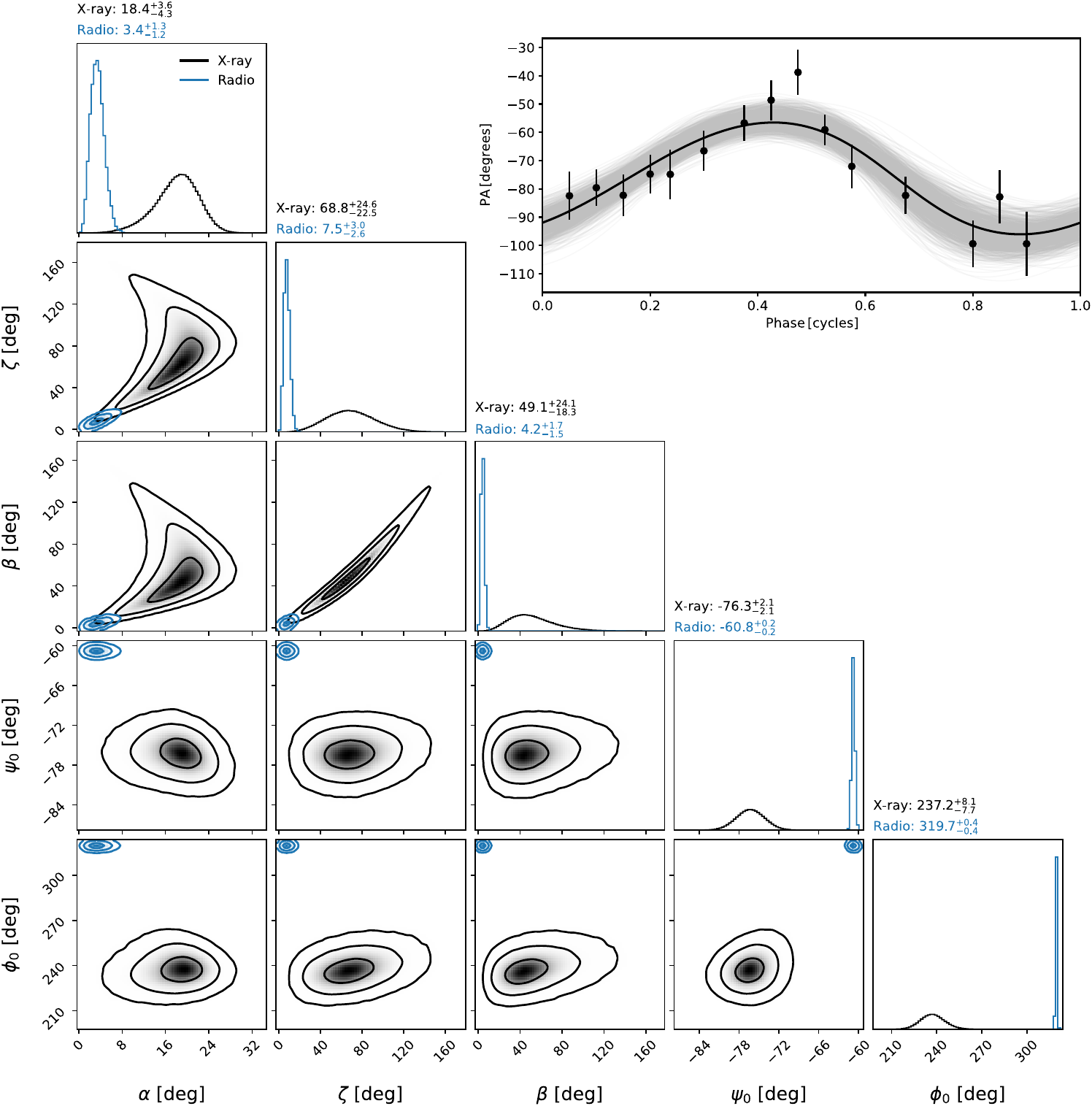}
\caption{RVM fit to the phase-resolved X-ray PA. {\sl Left panel}: Black contours represent the one- and two-dimensional posterior distributions for the RVM parameters $\alpha$, $\zeta$, $\beta$, $\psi_0$, and $\phi_0$, obtained from a Bayesian fit to the 2-4 keV phase-resolved polarization angle profile using the \cite{NKC1993AA:xrayrvm} wrapped-angle likelihood. Contours denote the 39\%, 87\%, and 99\% credible regions (corresponding to 1, 2, and 3$\sigma$). The blue contours are those derived from the radio RVM fit overlaid for ease of comparison. {\sl Right panel}: Phase-resolved X-ray PA measurements (black points) overlaid with the maximum-posterior RVM model (solid black line). Light gray curves show random posterior draws, illustrating the range of model realizations consistent with the data.}
	\label{fig:xrayrvmpost}
\end{figure}

\begin{figure}[H]
	\centering
	\includegraphics[width=0.95\linewidth]{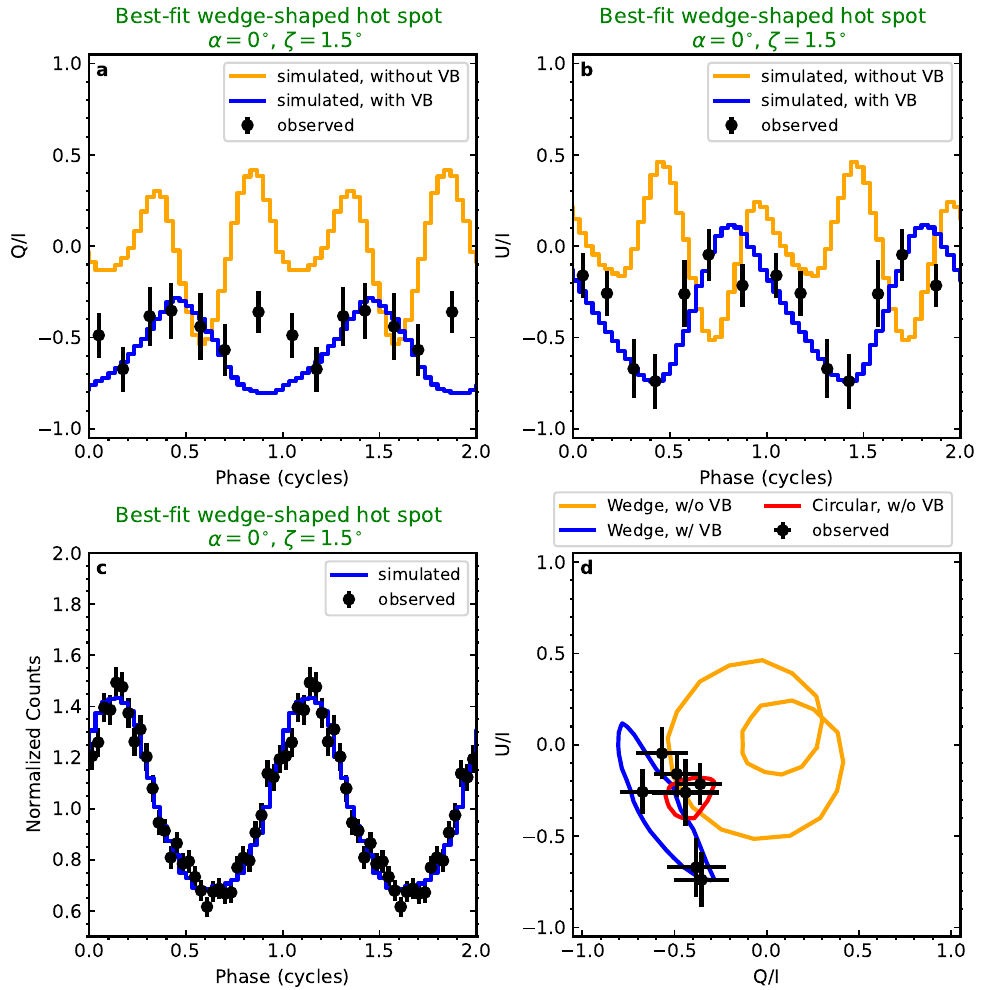}
	
	\caption{Simulated  Stokes $Q/I$ ({panel a}), $U/I$ ({panel b}),  and intensity ({panel c}) pulse profiles from {\sl MAGTHOMSCATT} for a single hot spot wedge offset from the magnetic pole as a function of rotational phase (solid lines). The black dots represent the  Stokes $Q/I$ ({panel a}),  $U/I$ ({panel b}), and intensity ({panel c}) data extracted from IXPE in the 2--3 keV energy range. The case with the best statistical fit (lowest combined total $\chi^2$; see Methods) incorporates vacuum birefringence, and corresponds to a magnetic colatitude of $\theta_m = [0\degree, 17\degree]$ and longitude $\phi_m = [0\degree, 120\degree]$ (zero longitude contains the rotation and magnetic axes); it is displayed in blue. The orange solid curves on the center and right panel show the corresponding polarization profiles wherein VB is turned off – these fits are statistically worse than those with VB on (see Methods). {Panel (d) displays the comparison in the Stokes Q-U space between the observed data (black dots) and the simulated results obtained with the best-fit wedge-shaped hot spot with (blue line) and without (orange line) including magnetospheric VB, plotted for one rotational cycle. The red solid line represents the best result among the VB-off cases (see Methods), corresponding to a pole-centered circular hot spot with a magnetic inclination of $\alpha = 2^{\degree}$ and a viewing angle of $\zeta = 20.5^{\degree}$.}}
	\label{fig:wedge-simulations}
\end{figure}

\begin{table}[h]
	\centering
	\caption{Timing and Pulsar Parameters}\label{tab:timing}
	%\begin{comment}
	
	\begin{tabular}{l l}
		\toprule
		\multicolumn{2}{l}{Pulsar parameters} \\
		\midrule R. A. (J2000) ${ }^{\text {a }}$ & $15^{\mathrm{h}} 50^{\mathrm{m}} 54.124$\\
		Decl. (J2000) ${ }^{\text {a }}$ & $-54\degree 18^{\prime} 24.11$ \\
		$\nu(\mathrm{Hz})$ & 0.477856(3) \\
		$\dot{\nu}\left(\mathrm{s}^{-2}\right)$ & $-4.(4) \times 10^{-12}$ \\
		Timing epoch (MJD TDB) & 58350 \\
		Position epoch (MJD TDB) & 54795 \\
		DM (pc cm$^{-3}$) & 697(1) \\
		Time span (MJD) & 60760-60770 \\
		Solar system ephemeris & DE405 \\
		{TZRMJD } & {60764.7495 } \\
		{TZRFRQ } & { 3275.71 } \\
		{TZRSITE } & { pks } \\
		\bottomrule
	\end{tabular}
	%\end{comment}
	\textbf{Extended Data Table 1: Timing and pulsar parameters.} Timing and pulsar parameters of the magnetar \src\ from  joint \textit{Murriyang} + NICER observations. Values in parentheses are the $1 \sigma$ uncertainties on the last digit. ${ }^{\mathrm{a}}$ Fixed to values from \cite{2012Deller}.
\end{table}

\begin{table}[h]
	\centering
	\caption{\textbf{Observations}}
	%\begin{comment}
	\begin{tabular}{c c c c c}
		\toprule
		Telescope & Observation ID & Start Date (UTC) & Stop Date (UTC) & Total Livetime GTI (ks) \\
		\midrule
		\midrule
		IXPE & 04003801 & 2025-03-26T03:01:52 & 2025-04-05T00:26:53 & 499.5 \\
		NICER & 8020300101 & 2025-03-26T01:09:00 & 2025-03-26T15:13:00 & 1.057 \\
		NICER & 8020300102 & 2025-03-27T01:51:01 & 2025-03-27T20:40:58 & 1.42\\
		NICER &8020300103 & 2025-03-28T14:56:37 & 2025-03-28T22:47:32 & 1.051 \\
		NICER & 8020300104 & 2025-03-31T06:33:41 & 2025-03-31T08:16:16 & 0.996 \\
		Murriyang & r070819\_124237 & 2007-08-19T12:42:37 & 2007-08-19T12:17:06 & 0.869 \\
		Murriyang & uwl\_250326\_173709 & 2025-03-26T17:37:09 & 2025-03-26T18:29:50 & 3.171 \\
		Murriyang & uwl\_250330\_171906 & 2025-03-30T17:19:06 & 2025-03-30T17:59:36 & 2.436 \\
		Murriyang & uwl\_250331\_131207 & 2025-03-31T13:12:07 & 2025-03-31T13:59:53 & 2.873 \\
		\bottomrule
	\end{tabular}
	%\end{comment}
	\textbf{Extended Data Table 2: Observations.} List of Parkes/Murriyang, IXPE, and NICER observation IDs, observation start and end times, and total livetimes used in this analysis.
	\label{tab:obsid}
\end{table}

\section{Supplementary Materials}\label{sec:supplementary}

\subsection{IXPE data reduction}

The Imaging X-ray Polarimetry Explorer (IXPE), launched in December 2021, is the first imaging X-ray polarimetry space observatory. IXPE consists of three identical telescopes sensitive to 2--8 keV with detector units (DU) containing a Gas Pixel Detector (GPD) that images the photoelectron tracks from incoming X-rays \cite{weisskopf22JATIS}. 
IXPE observed the magnetar \src\ between UTC 2025-03-26T03:01:52 (MJD 60760.12629630) and UTC 2025-04-05T00:26:53 (MJD 60770.01866898) for a total livetime duration of 499.5 ks (see Extended Table 2). We applied background rejection via the treatment recommended in \cite{DiMarcoCut} using the software \textit{filter\_background.py}. No significant solar flaring or bursting activity was detected during this observation. The total number of counts detected in the full IXPE band is 20943. We performed the solar-system barycenter correction of the data using the HEASoft version 6.35 task \texttt{barycorr} according to the source J2000 coordinates, R.A = 237.725517\degree, Dec = -54.306697\degree \cite{2012Deller} with the JPL planetary ephemeris DE405. We subsequently used the pulsar timing package, PINT \cite{luo19ascl:pint}, to assign phases to each photon according to a phase-coherent timing solution that incorporated both radio and X-ray time of arrivals (ToAs) (see Extended Table 1 for details of the full timing solution).

%to be at \textbf{J2000 coordinates (237.7305564\degree, -54.3048543\degree)}

After using DS9 to identify the peak of the centroid of the IXPE observation, we utilized the software \texttt{ixpeobssim} version 31.0.1 to perform data cleaning and extract high level data products, including spectra, lightcurves, and binned polarization cubes. The \texttt{ixpeobssim} task \texttt{xpradialprofiles} allowed us to examine the radial profiles and identify the extraction region size that optimized the data's signal-to-noise ratio. Based on the radial profiles, we selected a circular source region of 0.5' and an annular background with an inner radius of 2.5' and outer radius of 4.5'. For the energy and phase-resolved products, we select the instrument response functions (IRFs) as ixpe:obssim*:v13 from \texttt{ixpeobssim}’s pseudo-CALDB which provides the arf, rmf, PSF, vignetting functions, modulation factors, and modulation responses corresponding to the observation date \cite{baldini_2022_softX_ixpe}. For the spectral products, we generated the arf and mrf for each detector via the HEASoft tool \texttt{ixpecalcarf} and utilized the corresponding responses within Xspec during the spectral fitting. We verified that swapping the ixpeobssim arf and rmf curves with those built with HEASoft, or vice versa, returned consistent results.

The intensity X-ray pulse profiles are binned according to 32 equally spaced bins. The phase-resolved polarization degree and angle, on the other hand, were calculated according to 7 inhomogeneous phase bins to sample the morphology seen in both the radio and the X-ray intensity profiles. 

\subsection{NICER data reduction}

The Neutron star Interior Composition Explorer (NICER) is a non-imaging soft X-ray telescope aboard the International Space Station sensitive between 0.2--12 keV with an effective collecting area of 1900 cm$^{2}$ peaking at 1.5 keV. Its superior timing resolution of $<300$~nanoseconds makes NICER ideal for tracking magnetar temporal behavior. The data was extracted and cleaned using the HEASoft version 6.35 tool NICERDAS version 13a; the files were calibrated with NICER CALDB version 20240206. We applied standard filtering via the \texttt{nicerl2} task. Each photon time of arrival was barycenter corrected to the solar system according to the JPL solar-system ephemeris DE405 using the HEASoft tool \texttt{barycorr}. We generated the lightcurves and spectra using the task \texttt{nicerl3}, including the NICER spectral background model, SCORPEON which can be simultaneously fit with the source spectra to model sky and non-X-ray background contributions. 

\subsection{Parkes/{\it Murriyang} data reduction}

We conducted three observations of \src\ concurrent to the IXPE campaign using {\it Murriyang}, the 64 m Parkes radio telescope, with the Ultra-Wideband Low (UWL) receiver system \cite{2020PASA...37...12H}. These observations were collected as part of the long-term Parkes Magnetar Monitoring Programme (project ID: P885). We recorded full Stokes pulsar fold-mode data using the {\sc medusa} signal processor, where the voltage stream was coherently de-dispersed at 697\,pc\,cm$^{-3}$ to correct for signal dispersion in the interstellar medium and then folded at the rotation period of the magnetar with 1024 phase bins covering each rotation of the magnetar. The data were subsequently averaged in time to form 30\,s duration sub-integrations, each containing 3328 frequency channels covering the 704--4032\,MHz band of the UWL. The start and end times and duration of each observation are listed in Extended Table~2. 
We only made use of the data above 2.5\,GHz in this work, as multi-path propagation through the interstellar medium induces substantial scatter broadening of the pulse profile below this cutoff frequency, resulting in a smearing of the phase-dependent polarization properties \cite{camilo2008ApJ}. Radio-frequency interference from artificial sources were excised from the data by flagging and zero-weighting the contaminated frequency channels using the {\tt paz} and {\tt pazi} tools in {\tt psrchive} to first remove known persistently affected channels, and then manually flag any remaining interference \cite{2004PASA...21..302H, 2012AR&T....9..237V}. The data were then polarization and flux calibrated following the procedure outlined in Ref. \cite{2020ApJ...896L..37L} using the full receiver model implemented in the {\tt pac} tool of {\tt psrchive}. Prior to each magnetar observation, we conducted a short, 80\,s duration off-source pointing where a linearly polarized switched noise diode of known brightness was injected into the data stream. This was then compared to on/off observations of the calibration source PKS~B1934$-$638 to determine the absolute gain and phase corrections. The polarimetric response of the instrument as a function of parallactic angle was corrected for using a model derived from a rise-to-set track of the bright millisecond pulsar PSR~J0437$-$4725. We then corrected for the effects of Faraday rotation in the Stokes Q and U data due to propagation in the magnetized interstellar medium by applying the known rotation measure of $-1847.6$\,rad\,m$^{-2}$ using the {\tt pam} tool with the {\tt --aux\_rm} flag to de-rotate the data to infinite frequency.

\subsection{Phase-coherent timing analysis}

We obtained ToAs from the radio data by cross-correlating the individual frequency-averaged sub-integrations from each observation with a noise-free template via the Fourier-domain Monte Carlo method implemented in the {\tt pac} tool in {\tt psrchive}. The profile template was generated from a set of von Mises functions that were fit to a fully time/frequency averaged copy of the first Murriyang total intensity profile via the {\tt paas} tool. We derived X-ray ToAs from the IXPE and NICER data utilizing the Code for Rotational-analysis of Isolated Magnetars and Pulsars ({\tt CRIMP}). We create an X-ray pulse template from a high S/N NICER profile, which was then fit to a sinusoidal function including only the fundamental harmonic; adding more harmonics did not improve the quality of the fit. We then merge each instrument's event files, and split them into time intervals which incorporate 2000 counts and 4000 counts for NICER and IXPE, respectively, ensuring that each interval is $\lesssim0.5$~days. Finally, we fit the template to the phase-folded X-ray profiles of each interval, fixing all template parameters except for a phase-shift (i.e., the ToA) and the unpulsed component (which could show mild variability in NICER due to its non-negligible background).

An initial phase-coherent timing solution spanning the IXPE observing campaign was derived from the joint NICER and IXPE X-ray ToAs. We then used the tempo2 pulsar timing package \cite{2006MNRAS.369..655H, 2006MNRAS.372.1549E} to refine this solution by re-fitting the magnetar spin frequency and spin-down rate while including the radio ToAs, thereby obtaining a locally phase-coherent timing model covering both the radio and X-ray observations. All ToAs were barycentered using the JPL planetary ephemeris DE405. We verified that both epoch-to-epoch and intra-observation variations in the radio pulse profile had a negligible impact on the final timing precision.

To ensure accurate absolute phase alignment between the radio and X-ray data sets, the final timing solution presented in Table 1 was obtained by fitting only the radio ToAs. This ensures that the rotational parameters are properly referenced to TZRMJD, which defines the phase zero-point of the radio profile template used to derive the radio ToAs, along with the associated phase-defining parameters TZRFRQ and TZRSITE. The NICER and IXPE X-ray data were then folded using this final timing solution with PINT \cite{luo19ascl:pint}.

We aligned the three Murriyang observations in phase using the above timing solution, and then averaged all three together in time and frequency to form a single, average polarization profile. The total linear polarization was computed as the quadrature sum of the Stokes Q and U profiles as $L = \sqrt{Q^{2} + U^{2}}$, which was subsequently de-biased using the method outlined in \cite{2001ApJ...553..341E}.

\let\oldthebibliography=\thebibliography
\let\oldendthebibliography=\endthebibliography
\renewenvironment{thebibliography}[1]{
	\oldthebibliography{#1}
}{\oldendthebibliography}
%%%%%%%%%%%%%%%%%%%%%%%  reference for methods %%%%%%%%%%%%%%%

\section*{Declarations}
%Some journals require declarations to be submitted in a standardised format. Please check the Instructions for Authors of the journal to which you are submitting to see if you need to complete this section. If yes, your manuscript must contain the following sections under the heading `Declarations':

%\begin{itemize}
%\item Conflict of interest/Competing interests (check journal-specific guidelines for which heading to use)
%\item Ethics approval and consent to participate
%\item Consent for publication
%\item Data availability 
%\end{itemize}

%\noindent
%If any of the sections are not relevant to your manuscript, please include the heading and write `Not applicable' for that section. 
\bmhead{Data availability}
NICER observations (ObsIDs: 8020300101, 8020300102, 8020300103, 8020300104) and IXPE observations (04003801) are readily accessible on the HEASARC data archive: \url{https://heasarc.gsfc.nasa.gov/W3Browse} (DOI: \url{10.25504/FAIRsharing.979d22}). Radio observations made by \textit{Murriyang/Parkes} (ObsIDs: r070819\_124237, uwl\_250326\_173709, uwl\_250330\_171906, \& uwl\_250331\_131207; DOI: \url{10.4225/08/52292AE9B2D80}, \url{10.25919/rzdr-pw25}, \& \url{10.25919/v5hn-4v34}) are publicly available from the CSIRO Data Access Portal (\url{https://data.csiro.au/}) following an 18 month proprietary period starting on the observation date.
%Source data for Figure(s) [number(s)] are provided with the paper. 

\bmhead{Code availability}
Data reduction and analysis of X-ray products were performed using publicly available software Heasoft version 6.35.0 (\url{https://heasarc.gsfc.nasa.gov/docs/software/lheasoft/}) from the High Energy Astrophysics Science Archive Research Center (HEASARC), particularly -- FTOOLs v. 6.35.1, SAOImage DS9 v. 8.4b1, and Xspec v. 12.15.0. Generation and calibration of the NICER event lists was also performed by HEASoft's NICERDAS version 12. The simulation and analysis framework ixpeobssim v. 31.1.0 was used to generate high-level IXPE data products: \url{https://ixpeobssim.readthedocs.io/en/latest/}. Additionally, the software filterbackground.py was used for treatment of the IXPE background, found here: \url{https://github.com/aledimarco/IXPE-background}. Timing analysis was performed using tempo2 (\url{https://github.com/mattpitkin/tempo2}), PINT (\url{https://github.com/nanograv/PINT}), and CRIMP (\url{https://github.com/georgeyounes/CRIMP/tree/main}). PyXspecCorner (\url{https://github.com/garciafederico/pyXspecCorner}) and corner.py (\url{https://corner.readthedocs.io/en/latest/}) was used to generate the X-ray spectro-polarimetric and radio RVM corner plots, respectively. Additional custom code for generating figures and performing analysis are contained within a Github repository linked here: \url{https://github.com/rae-stewart/Polarimetric-Analysis-of-1E-1547.0-5408}. Custom code for the {\sl MAGTHOMSCATT} Monte Carlo simulation are available upon reasonable request.

\bmhead{Acknowledgments}

This work reports observations obtained with the Imaging X-ray Polarimetry Explorer (IXPE), a joint US (NASA) and Italian (ASI) mission, led by Marshall Space Flight Center (MSFC). The research uses data products provided by the IXPE Science Operations Center (MSFC), using algorithms developed by the IXPE Collaboration (MSFC, Istituto Nazionale di Astrofisica - INAF, Istituto Nazionale di Fisica Nucleare - INFN, ASI Space Science Data Center - SSDC), and distributed by the High-Energy Astrophysics Science Archive Research Center (HEASARC). Murriyang, CSIRO’s Parkes radio telescope, is part of the Australia Telescope National Facility (\href{https://ror.org/05qajvd42}{https://ror.org/05qajvd42}) which is funded by the Australian Government for operation as a National Facility managed by CSIRO. We acknowledge the Wiradjuri people as the Traditional Owners of the Observatory site. This project was supported by resources and expertise provided by CSIRO IMT Scientific Computing, and made use of the Ngarrgu Tindebeek supercomputer at the OzSTAR national facility at Swinburne University of Technology. The OzSTAR program receives funding in part from the Astronomy National Collaborative Research Infrastructure Strategy (NCRIS) allocation provided by the Australian Government, and from the Victorian Higher Education State Investment Fund (VHESIF) provided by the Victorian Government. G.Y. acknowledges constructive discussion with Alexander Philippov on radio emission from magnetars, and Paul Ray on radio/X-ray timing analysis.
\bmhead{Funding Statement}
 The material is based upon work supported by NASA under award number 80GSFC24M0006. G.Y. acknowledges NASA support under grants 80NSSC25K7257 and 80NSSC25K0283, through which R.E.S. and A.V.K. are partially supported. M.E.L. is supported by an Australian Research Council (ARC) Discovery Early Career Research Award DE250100508. M.G.B. thanks NASA for generous support under grants 80NSSC24K0589, 80NSSC25K7257 and 80NSSC25K0079. W.C.G.H. acknowledges support through grant 80NSSC23K0078 from NASA. J.B.C. acknowledges support under NASA award number 80GSFC21M0006. F.C., A.K.H., T.E., C.P.H, P.K.,  M.N., P.S., and Z.W do not declare relevant funding.

\bmhead{Author's Contributions}
R.E.S. performed the X-ray polarization and spectral data analysis and contributed to writing the paper. H.D.T. and M.G.B. led the simulation analysis and theoretical interpretations and contributed to writing the paper. G.Y. is the principal investigator of the IXPE observation presented in this work, which was obtained through the NASA IXPE Guest Observer cycle 2 program. G.Y. performed the X-ray timing analysis and contributed to writing the paper. M.E.L. performed the radio observations and analysis, and contributed to writing the paper. M.N. provided support with the X-ray polarization analysis and contributed to writing the paper. F.C., J.B.C., T.E., A.K.H.,  W.C.G.H., C.P.H, P.K., P.S., A.V.K., and Z.W. provided comments and contributed to the writing of the paper. F.C. is the principal investigator of the Parkes P885 project.

\bmhead{Conflicts of interest}
The authors declare no conflict of interest. 

\bmhead{Additional Statements}
Supplementary Information is available for this paper. Correspondence and requests for materials should be addressed to Rachael Stewart (raestewart@gwu.edu) and Hoa Dinh Thi (hoa.dinh@rice.edu). Reprints and permissions information is available at www.nature.com/reprints. We grant to the Federal Government, a royalty-free, nonexclusive and irrevocable license to use, reproduce, distribute (including distribution by transmission) to the public, perform publicly, prepare derivative works, and display publicly, data in whole or in part and in any manner for Federal purposes and to have or permit others to do so for Federal purposes only.

%TC:endignore
\end{document}